\documentclass[12pt,english,round]{article}
\usepackage[T1]{fontenc}
\usepackage{textcomp}
\usepackage[latin9]{inputenc}
\usepackage{color}
\usepackage{url}
\usepackage{amsmath}
\usepackage{amssymb}
\usepackage{geometry}
\usepackage{setspace}
\usepackage[authoryear]{natbib}
\usepackage{babel}
\begin{document}
\title{Modeling Uncertainty in Integrated Assessment Models\vspace{1cm}
}
\author{Yongyang Cai\thanks{Department of Agricultural, Environmental and Development Economics,
The Ohio State University, 2120 Fyffe Road, Columbus, OH, USA 43210.
cai.619@osu.edu.}}
\maketitle
\begin{abstract}
Integrated Assessment Models (IAMs) are pivotal tools that synthesize
knowledge from climate science, economics, and policy to evaluate
the interactions between human activities and the climate system.
They serve as essential instruments for policymakers, providing insights
into the potential outcomes of various climate policies and strategies.
Given the complexity and inherent uncertainties in both the climate
system and socio-economic processes, understanding and effectively
managing uncertainty within IAMs is crucial for robust climate policy
development. This review aims to provide a comprehensive overview
of how IAMs handle uncertainty, highlighting recent methodological
advancements and their implications for climate policy. I examine
the types of uncertainties present in IAMs, discuss various modeling
approaches to address these uncertainties, and explore recent developments
in the field, including the incorporation of advanced computational
methods.



\newpage{}
\end{abstract}

\section{Introduction}

Integrated assessment models (IAMs) serve as crucial tools at the
intersection of climate science, economics, and policy. They synthesize
knowledge across disciplines to evaluate the interactions between
the climate system and socio-economic activities, providing insights
into mitigation pathways, adaptation strategies, and long-term risks.
These models enable cost-benefit assessments of climate policies,
the computation of the Social Cost of Carbon (SCC), and exploration
of technological and policy pathways under various socio-economic
and environmental scenarios. IAMs have become central to international
climate policy, particularly through their influence on reports by
the Intergovernmental Panel on Climate Change (IPCC). However, IAMs
are fundamentally challenged by the presence of uncertainty---ranging
from scientific ambiguity and socioeconomic variability to risk and
normative uncertainty about future preferences. Addressing these uncertainties
in a structured manner is essential for producing credible policy-relevant
insights.

Uncertainty permeates every layer of IAMs, from physical climate sensitivity
and risks to economic risks, climate damage functions and future policy
responses. Uncertainty in integrated assessment modeling can be categorized
in various ways. The most prominent typologies distinguish among parameter
uncertainty, scenario uncertainty, model or structural uncertainty,
risk, and deep uncertainties. Each of these has significant implications
for the robustness and reliability of IAM outputs. Decisions made
under such uncertainty have profound implications for climate action,
risk management, and equity. Recent years have seen substantial advancements
in how IAMs incorporate uncertainty. This includes stochastic modeling
techniques, applications of Epstein-Zin preferences to separate risk
aversion and intertemporal elasticity of substitution, robust control
approaches to ambiguity, and learning models that account for evolving
knowledge. Machine learning has also emerged as a promising tool to
enhance the flexibility and accuracy of IAMs. Except modeling uncertainty
directly in IAMs, we can also have a more detailed representation
of a system to reduce uncertainty, such as disaggregation of space,
sectors, and agents. This review explores how IAMs incorporate and
respond to uncertainty, highlighting methodological innovations, and
evaluating the implications for policy design and robustness. Because
of space limitations, this review focuses on recent developments of
IAMs, and skips those reduced-form IAMs.\footnote{See van der Ploeg and Zeeuw \citeyearpar{van_der_ploeg_climate_2018,van_der_ploeg_pricing_2019},
\citet{van_der_ploeg_risk_2020}, \citet{HAMBEL_EER2021}, \citet{iverson2021carbon},
\citet{Bremer_AER2021}, \citet{fowlie2022mitigating}, \citet{goulder2022china},
\citet{Olijslagers2024}, and \citet{Zhu_2025}, for some recent examples
of reduced-form IAMs. } Readers can consult \citet{LemoineRudik2017}, \citet{Bosetti2021},
\citet{Cai2021}, \citet{bohringer_NCC_2022}, \citet{Desmet2024},
\citet{DIETZ2024}, \citet{Bilal2025}, and \citet{Fernandez-Villaverde2025}
for more reviews about IAMs. 

The rest of the article is organized as follows. It begins with a
discussion of the most popular IAM, the Dynamic Integrated model of
Climate and the Economy model (DICE), developed by Professor William
Nordhaus. This discussion is followed by a review of extensions of
DICE in Section 3, other key IAMs in Section 4, and computational
methods for handling uncertainty in Section 5. Finally summary points
and future issues are presented.

\section{DICE}

DICE integrates a neoclassical economic growth model with a simplified
climate system. It has been instrumental in estimating the social
cost of carbon (SCC) and informing optimal carbon pricing strategies.
DICE has been updated with many versions since its first development
in 1992 \citep{Nordhaus1992}. This review uses DICE-2016 \citep{Nordhaus_DICE2016}
as an illustrative example of IAMs. 

DICE-2016 is a dynamic programming model for maximizing the global
social welfare under infinite time horizon, where the time step size
is five years while it is one decade in the earlier versions of DICE
like DICE-2007 \citep{Nordhaus_DICE2007}.\footnote{\citet{cai_open_2012} show that the decadal time step size increases
the optimal carbon tax up to roughly 50\% higher, compared with a
solution of an annualized DICE version. } It has six endogenous state variables: capital $K_{t}$ in the economic
system, three layers of carbon concentrations $\mathbf{M}_{t}=(M_{\mathrm{AT},t},M_{\mathrm{UO},t},M_{\mathrm{DO},t})'$
(in the atmosphere, upper ocean, and deep ocean) in the carbon cycle,
and two layers of temperatures $\mathbf{T}_{t}=(T_{\mathrm{AT},t},T_{\mathrm{OC},t})'$
(atmospheric and oceanic) measured as temperature increases above
the preindustrial levels. There are two decision variables at each
period: consumption $C_{t}$ for social utility, and emission control
rate $\mu_{t}$ for mitigation of emissions. 

DICE assumes that climate damage is proportional to gross production
$Y_{t}=A_{t}K_{t}^{\alpha}L_{t}^{1-\alpha}$, where $A_{t}$ is exogenous
total factor of productivity and $L_{t}$ is exogenous global population
size at time $t$, and the damage proportion is $1-\Omega\left(T_{\mathrm{AT},t}\right)$
with
\[
\Omega\left(T_{\mathrm{AT},t}\right)=\frac{1}{1+\pi_{1}T_{\mathrm{AT},t}+\pi_{2}(T_{\mathrm{AT},t})^{2}}
\]
where $\pi_{1}$ and $\pi_{2}$ are parameters. The mitigation expenditure
$\Psi_{t}$ is also assumed to be proportional to gross production:
$\Psi_{t}=\theta_{1,t}\mu_{t}^{\theta_{2}}Y_{t}$, where {\small\textcolor{black}{$\theta_{1,t}$
is }}exogenous{\small\textcolor{black}{{} adjusted cost for backstop}},
and $\theta_{2}$ is a parameter. Total carbon emissions are 
\begin{equation}
E_{t}=\sigma_{t}(1-\mu_{t})Y_{t}+E_{\mathrm{Land},t},\label{eq:DICE-E}
\end{equation}
where $\sigma_{t}$ is exogenous carbon intensity of output, and $E_{\mathrm{Land},t}$
represents exogenous land emissions. 

DICE solves the following dynamic programming model: 
\begin{eqnarray*}
\max_{C_{t},\mu_{t}} &  & \sum_{t=0}^{\infty}\beta^{t}u(C_{t},L_{t})
\end{eqnarray*}
subject to transition laws of the six state variables: 
\begin{eqnarray}
K_{t+1} & = & (1-\delta)K_{t}+\widehat{Y}_{t}-C_{t}\label{eq:DICE-K}\\
\mathbf{M}_{t+1} & = & \Phi_{M}\mathbf{M}_{t}+\left(E_{t},0,0\right)^{\top}\label{eq:DICE-M}\\
\mathbf{T}_{t+1} & = & \Phi_{T}\mathbf{T}_{t}+\left(\xi_{1}\mathcal{F}_{t}\left(M_{\mathrm{AT},t}\right),0\right)^{\top}\nonumber 
\end{eqnarray}
Here $\beta$ is discount factor, 
\[
u(C_{t},L_{t})=\frac{(C_{t}/L_{t})^{1-\frac{1}{\psi}}}{1-\frac{1}{\psi}}L_{t}
\]
is the utility function with $\psi$ being inter-temporal elasticity
of substitution, $\delta$ is depreciation rate of capital, 
\[
\widehat{Y}_{t}=\Omega\left(T_{\mathrm{AT},t}\right)Y_{t}-\Psi_{t}
\]
is output net of climate damage and mitigation expenditure, $\Phi_{M}$
is a $3\times3$ matrix, $\Phi_{T}$ is a $2\times2$ matrix, $\xi_{1}$
is a parameter related to equilibrium climate sensitivity (i.e., the
long-run increase of surface temperature in °C from a doubling of
carbon concentration in the atmosphere), and 
\[
\mathcal{F}_{t}\left(M_{\mathrm{AT},t}\right)=\eta\log_{2}\left(\frac{M_{\mathrm{AT},t}}{M_{\mathrm{AT}}^{*}}\right)+F_{\mathrm{EX},t}
\]
where $M_{\mathrm{AT}}^{*}$ is the pre-industrial atmospheric carbon
concentration, $\eta$ is a parameter, and $F_{\mathrm{EX},t}$ is
exogenous radiative forcing. 

For climate policy analysis, DICE is thus used for calculating the
SCC, which is the present value of future additional climate damage
caused by one additional unit of carbon emissions at the current period.
In DICE, the SCC is defined as the marginal rate of substitution in
the social welfare between global carbon emissions and aggregate consumption,
which is numerically calculated as the ratio of the shadow price of
the equation of global emissions (\ref{eq:DICE-E}) to the shadow
price of the capital transition equation (\ref{eq:DICE-K}). The optimal
carbon tax is computed via $\theta_{1,t}\theta_{2}\mu_{t}^{\theta_{2}-1}/\sigma_{t}$,
which should be equal to the SCC according to a Pigovian tax policy
when $\mu_{t}$ is not binding at its bounds. 

However, there is great uncertainty in DICE. The values of the parameters
and the exogenous time paths in DICE are estimated from historical
data or projections for future scenarios. DICE is designed for estimating
climate impact and policy in future, but future realized data could
be beyond the range of historical data and any projections for future
scenarios may be not close to what will actually happen, therefore
the values of the parameters and the projected time paths are uncertain.
These can be classified as parameter uncertainty\footnote{In this review, parameter uncertainty is used to represent the cases
in which an uncertain parameter has an unknown true value that is
unchanged over time, so it can be distinguished with risk \citep{Cai2021}. } and scenario uncertainty respectively.

For example, \citet{Stern2007} argues a discount factor $\beta=0.999$
for ethical reasons (while DICE assumes $\beta=0.985$) and finds
that the SCC will be significantly higher. Interagency Working Group
on Social Cost of Carbon employs three social discount rates (2.5\%,
3\%, and 5\%), which are connected with $\beta$ and $\psi$ by the
famous Ramsey rule, to compute the SCC. \citet{drupp_discounting_2018}
conduct an online survey to potential experts, and find that the median,
mean, and standard deviation of social discount rates are 2, 2.27,
and 1.62, respectively, and that the median, mean, and standard deviation
of elasticity of marginal utility (i.e., $1/\psi$) are 1, 1.35, and
0.85, respectively, while DICE-2007 sets the elasticity of marginal
utility to $2$ and DICE-2016 changes it to $1.45$. Equilibrium climate
sensitivity is a well-known uncertain parameter in the climate system.
\citet{IPCC2021} suggests the best estimate of equilibrium climate
sensitivity is 3°C, the likely range (i.e. with a 66\% probability)
is $[2.5,4.0]$. The climate damage estimation is also debated extensively.
For example, \citet{Weitzman2012} suggests adding one high-exponent
term to the quadratic function in $\Omega\left(T_{\mathrm{AT},t}\right)$
so that 50\% of output is lost if the temperature increase is 6 °C,
to avoid implausibly low damage at high temperatures in $\Omega\left(T_{\mathrm{AT},t}\right)$
used in DICE. This modification will lead to a significantly higher
SCC \citep{DietzStern2015}. In addition, \citet{Meinshausen_RCP}
describe four Representative Concentration Pathways (RCPs) of greenhouse
gas concentrations: RCP2.6, RCP4.5, RCP6, and RCP8.5, which are often
used for calibrating the parameters in the climate system (e.g., \citet{CaiJuddLontzek2017_DSICE,CaiLontzek2019_DSICE}).
\citet{Folini2025} recalibrate the climate system of DICE-2016 with
benchmark data from comprehensive global climate models in the Coupled
Model Intercomparison Project, Phase 5 (CMIP5, \citet{navarro-racines_2020}),
and find that the climate system of DICE-2016 are mis-calibrated,
implying uncertainty in the parameters in the matrices $\Phi_{M}$
and $\Phi_{T}$. 

Except the parameters' uncertainty, the exogenous time paths in DICE
are also uncertain. For example, \citet{ONeill_etal2014} provide
five Shared Socio-Economic Pathways (SSPs): SSP1 (Sustainability),
SSP2 (Middle of the Road), SSP3 (Regional Rivalry), SSP4 (Inequality),
and SSP5 (Fossil-fueled Development), which cover wide ranges of the
projected time paths of population, income, and temperature, based
on different assumptions about efforts toward sustainability and socioeconomic
development goals. That is, each SSP scenario has its associated population
$L_{t}$, total factor of productivity $A_{t}$, carbon intensity
$\sigma_{t}$, and so on. 

Except the scenario uncertainty, DICE itself has also model or structural
uncertainty. For example, DICE-2023 \citep{BarrageNordhaus2023},
the most recent version of DICE, replaces the climate system of DICE-2016
by D-FAIR: the DICE version of the FAIR (Finite Amplitude Impulse-Response)
model \citep{millar_FAIR_2017} including four reservoirs for carbon
concentration and two temperature boxes. This implies that the model
structure of DICE is also uncertain. 

\section{Extensions of DICE}

The last section discusses parameter uncertainty, scenario uncertainty,
and model or structural uncertainty in DICE, but DICE itself is a
deterministic model, that is, it also ignores risks in the economic
and climate systems. Here risks refer to random variables with probability
distributions that are known or have known functions of state or control
variables at each time period \citep{Cai2021}. That is, DICE can
be extended to be stochastic to deal with risk aversion. Moreover,
DICE can also be extended with a more detailed representation of the
economic and climate systems to reduce model or structural uncertainty,
which arises from the limitations or simplifications in the model\textquoteright s
representation of complex systems. With disaggregation over space,
sectors, and agents, we can then discuss and compare various climate
policies---such as (regional) tax, subsidy, and cap-and-trade---and
their associated policy analysis under cooperation or noncooperation.
There are many extensions of DICE, here I discuss only some recent
extensions due to space limitations. 

\subsection{Stochastic Extension }

\citet{CaiJuddLontzek2017_DSICE} and \citet{CaiLontzek2019_DSICE}
extend the full DICE-2007 model \citep{Nordhaus_DICE2007} to a dynamic
stochastic framework, called DSICE (Dynamic Stochastic Integrated
framework of Climate and Economy), incorporating long-run economic
risk, climate tipping risk, and Epstein-Zin preferences \citep{Epstein-Zin-1989}.
Because the time-separable utility in DICE does not explain the willingness
of people to pay to avoid risk, DSICE uses Epstein--Zin preferences
to explain equity or insurance premia about how much society is willing
to pay to reduce the risk of economic damage from climate change.
That is, in DSICE a social planner maximizes the following recursively
defined social welfare \textcolor{black}{
\begin{equation}
U_{t}=\Xi\left\{ \left(1-\beta\right)\Xi u(C_{t},L_{t})+\beta\left[\mathbb{E}_{t}\left\{ \left(\Xi U_{t+1}\right)^{1-\gamma}\right\} \right]^{\frac{1}{\varTheta}}\right\} ^{\frac{1}{1-1/\psi}},\label{eq:EZ-U}
\end{equation}
where $\psi$ is the intertemporal elasticity of substitution, $\gamma$
is the risk aversion} coefficient,\textcolor{black}{{} }$\varTheta=(1-\gamma)/(1-1/\psi)$,\textcolor{black}{{}
$\Xi=\mathrm{sgn}(\psi-1)$ is the sign function of $\psi-1$ (that
is, $\Xi=1$ if $\psi>1$, or $-1$ otherwise), and $\mathbb{E}_{t}\mathbb{\left\{ \cdot\right\} }$
is the expectation conditional on the states at time $t$.}

In the economic system, DSICE replaces the DICE's exogenous deterministic
total factor of productivity $A_{t}$ by$A_{t}\zeta_{t}$, with a
shock $\zeta_{t}$ following a dense Markov chain discretized from
the following long-run risk process: \textcolor{black}{
\begin{equation}
\log\left(\zeta_{t+1}\right)=\log\left(\zeta_{t}\right)+\chi_{t}+\varrho\omega_{\zeta,t}\label{eq:zeta_process}
\end{equation}
\begin{equation}
\chi_{t+1}=r\chi_{t}+\varsigma\omega_{\chi,t},\label{eq:chi_process}
\end{equation}
where $\chi_{t}$ represents the stochastic persistence of the shock
}$\zeta_{t}$\textcolor{black}{, $\omega_{\zeta,t},\omega_{\chi,t}\sim i.i.d.\:\mathcal{N}(0,1)$,
and $\varrho$, $r$, and $\varsigma$ are parameters.} The discretization
of $\zeta_{t}$ and \textcolor{black}{$\chi_{t}$} changes the unbounded
normal distributions of $\omega_{\zeta,t}$ and $\omega_{\chi,t}$
to be bounded such that the expectations in (\ref{eq:EZ-U}) are finite. 

In the climate system, DSICE considers climate tipping risk, which
refers to a probable transition to an irreversible state of the climate
system. \textcolor{black}{Examples of climate tipping risks include}
Atlantic meridional overturning circulation, disintegration of the
Greenland ice sheet, and collapse of the West Antarctic ice sheet\textcolor{black}{.
}In the benchmark tipping examples, DSICE replaces the DICE's climate
damage factor $\Omega\left(T_{\mathrm{AT},t}\right)$ by 
\[
\Omega\left(T_{\mathrm{AT},t},J_{t}\right)=\frac{1-J_{t}}{1+\pi_{1}T_{\mathrm{AT},t}+\pi_{2}(T_{\mathrm{AT},t})^{2}}
\]
where $J_{t}$ is a Markov chain representing an irreversible climate
tipping process with 16 possible values of tipping damage levels $\left\{ \mathcal{J}_{1},\mathcal{J}_{2},...,\mathcal{J}_{16}\right\} $,
where $\mathcal{J}_{1}=0$ represents the pre-tipping stage. The tipping
probability is 
\[
p_{\mathrm{tip},t}=1-\exp\left\{ -\lambda\max\left\{ 0,\,T_{\mathrm{AT},t}-\underline{T_{\mathrm{AT}}}\right\} \right\} 
\]
where $\lambda$ is the hazard rate parameter, and $\underline{T_{\mathrm{AT}}}$
is the temperature threshold without tipping. Except the uncertainty
of tipping time, DSICE also assumes the duration of the tipping process
is uncertain and the final damage level from tipping is uncertain
too. The climate tipping process in the benchmark tipping examples
of DSICE can be divided into the pre-stage stage, four transient stages,
and the final absorbing stage. The duration of each transient stage
is assumed to follow an exponential distribution with mean $\overline{\Gamma}/4$.
The long-run damage level at the final absorbing stage, denoted $\mathcal{J}_{\infty}$,
is assumed to be stochastic with mean $\overline{\mathfrak{D}}_{\infty}$
and variance $q\overline{\mathfrak{D}}_{\infty}^{2}$. These model
the gradual nature of the tipping process and uncertainty about the
ultimate damage caused. For convenience, \textcolor{black}{the tipping
process is denoted} as 
\[
J_{t+1}=g_{J}(J_{t},\mathbf{T}_{t},\omega_{J,t}),
\]
\textcolor{black}{where $\omega_{J,t}$ is a serially independent
stochastic process. }

After the DSICE model is set up, we can numerically solve it, then
do economic and policy analysis, such as the calculation of the SCC
and the optimal carbon tax. The SCC in DSICE is computed as the marginal
rate of substitution in the expected social welfare between atmospheric
carbon concentration and capital, that is, 
\[
\mathrm{SCC}_{t}=-\left(\partial V_{t}/\partial M_{\mathrm{AT},t}\right)/\left(\partial V_{t}/\partial K_{t}\right)
\]
 where $V_{t}$ represents the optimal expected social welfare at
the state vector $\left(K_{t},\mathbf{M}_{t},\mathbf{T}_{t},\zeta_{t},\chi_{t},J_{t}\right)$
at time $t$. The optimal carbon tax has the same form with DICE,
i.e., $\theta_{1,t}\theta_{2}\mu_{t}^{\theta_{2}-1}/\sigma_{t}$. 

Under the above specified DSICE framework, the benchmark stochastic
growth examples in \citet{CaiJuddLontzek2017_DSICE} and \citet{CaiLontzek2019_DSICE}
show that the long-run growth risk leads to a stochastic process of
the SCC with a wide range of possible values, and the recursive utility's
preference parameters have a nontrivial impact on the SCC: with a
large inter-temporal elasticity of substitution, a larger risk aversion
implies a smaller SCC; with a small inter-temporal elasticity of substitution,
a larger risk aversion implies a larger SCC. The benchmark tipping
examples in \citet{CaiJuddLontzek2017_DSICE} and \textcolor{black}{\citet{CaiLontzek2019_DSICE}
show that a higher }inter-temporal elasticity of substitution\textcolor{black}{{}
or risk aversion always leads to a higher SCC. If a tipping event
has not happened, then the SCC is significantly higher than in DICE,
because of the incentive to }prevent or delay the tipping event.\textcolor{black}{{}
But once the tipping event happens, the SCC will jump down significantly
and immediately as the incentive has disappeared, even though the
post-tipping damage has a long duration to reach its long-run damage
level. The benchmark examples with both stochastic growth and climate
risks in }\citet{CaiJuddLontzek2017_DSICE} and \citet{CaiLontzek2019_DSICE}
show that the interaction between the long-run risk and the tipping
process has nontrivial impacts to results: for example, while either
the long-run risk or the tipping process leads to a higher SCC than
in DICE, their combination does not imply a further increase in the
SCC when compared to the cases with only one type of risk. All the
examples in \citet{CaiJuddLontzek2017_DSICE} and \citet{CaiLontzek2019_DSICE}
show that risks can have significant impact on the SCC, and those
reduced form IAMs could lead to misleading results. For example, \citet{Golosov2014}
use a reduced form IAM with logarithmic utility and full capital depreciation
to argue that the SCC is proportional to output with a constant ratio,
but every numerical example in \citet{CaiJuddLontzek2017_DSICE} and
\citet{CaiLontzek2019_DSICE} shows that the ratio is stochastic and
its variance is not small.

The DSICE framework has also been applied with various variants in
\citet{lontzek_NCC_2015}, \citet{Cai_PNAS2015}, and \citet{cai_NCC_2016}.
\citet{lontzek_NCC_2015} model a climate tipping risk with a continuous
tipping damage path, and find that the optimal carbon tax increases
significantly even with conservative assumptions about the rate and
impacts of a stochastic tipping event.\footnote{See \citet{Dietz_PNAS_2021} and \citet{Armstrong_McKay_Science2022}
for discussion of various climate tipping risks. } \citet{Cai_PNAS2015} model an environmental tipping risk with climate
damages on market and nonmarket goods and services, and find that
the nonmarket impacts could substantially increase the optimal carbon
tax. \citet{cai_NCC_2016} consider five major interacting climate
tipping risks (Atlantic meridional overturning circulation, disintegration
of the Greenland ice sheet,\footnote{\citet{nordhaus_economics_2019} finds that the risk of Greenland
ice sheet disintegration makes a small contribution to the overall
social cost of climate change, by modeling the risk as a deterministic
and endogenous process. It is consistent with \citet{cai_NCC_2016}
in the case without interactions between tipping events.} collapse of the West Antarctic ice sheet, dieback of the Amazon rainforest,
and shift to a more persistent El Niño regime), and find that these
increase the initial SCC by nearly eightfold. Moreover, passing a
tipping point could abruptly increase the SCC if it increases the
likelihood of other tipping events. This incorporation of tipping
elements reflects an important evolution of IAMs toward modeling catastrophic
or low-probability, high-impact events, which are often underrepresented
in earlier generations of models.

\subsection{Spatial Disaggregation for Economy}

DICE models the global economy with only one global capital and one
global production function, but ignoring regional heterogeneity in
the economic system leads to large uncertainty in estimation for regional
and country-level economic and climate policies. In particular, it
is challenging to impose a global carbon tax to every country. The
Regional Integrated model of Climate and the Economy (RICE) extends
DICE by spatially disaggregating the world into multiple regions in
the economic system and solves cooperative equilibria with weights
on regional utilities. It has three versions: RICE-1996 \citep{nordhaus_yang_1996}
with six regions, RICE-2010 \citep{Nordhaus_RICE_2010} with 12 regions,
and RICE-2020 \citet{yang_model_2023} with a flexible number of regions
up to 16. Moreover, RICE-1996 and RICE-2020 study non-cooperative
equilibria with open-loop Nash equilibrium solutions, assuming that
the regions are noncooperative and maximize their own utility only
taking into account climate change damages to their own output. An
open-loop Nash equilibrium solution depends on only the initial condition
and time, and it could be interpreted as a situation in which the
regions enter an agreement to commit to a future path of carbon emissions
at the beginning of the agreement. The cooperative and noncooperative
solutions are two extreme cases that can provide helpful analysis
to policymakers. 

\citet{CaiMalikShin2023} extend RICE to incorporate a global emission
trading system within 12 world regions, assuming every country in
the regions will follow their commitments in nationally determined
contributions under the Paris Agreement and the Glasgow Climate Pact.
An emission trading system, also known as a cap-and-trade scheme,
fixes the maximum amount of emission allowances for the market to
trade, so it provides direct control over future emissions and it
would be more straightforward to control temperature increase under
some threshold. \citet{CaiMalikShin2023} also replace the DICE climate
system by a simpler but more stable climate system called transient
climate response to emissions \citep{matthews_proportionality_2009},
which assumes that contemporaneous globally average atmospheric temperature
increase is linearly proportional to cumulative global carbon emissions.
They calculate the endogenous emission permit prices under an open-loop
Nash equilibrium solution with a competitive equilibrium in the emission
permit market, without assuming the equality of marginal abatement
costs across regions. They demonstrate that the regional SCC is the
difference between regional marginal abatement cost and the global
permit price, both theoretically and numerically, implying the complementarity
between carbon tax and emission trading system. 

\subsection{Spatial Disaggregation for Climate}

The spatial disaggregation of RICE follows political and legal jurisdictions,
but its climate system still follows DICE by using the globally averaged
measure of temperature, which ignores heterogeneity in the regional
temperatures, especially polar amplification, which means that warming
in the high latitudes increases faster than in the tropical region.
To address this, \citet{cai2023_JAERE} partition the globe into three
regions by following physical laws in modeling the regional climate
systems (i.e., heat and moisture transfer between regions). The three
regions are the North, the Tropics, and the South, and heat and moisture
transport are from the Tropics to the other regions. They use the
four RCP scenarios of emissions and atmospheric carbon concentration
to calibrate the parameters in the matrix $\Phi_{M}$ in the transition
equation (\ref{eq:DICE-M}) for the carbon cycle system, and the ensemble
mean of CMIP5 models' annual projections of temperature anomaly in
every region under the four RCP scenarios to calibrate the parameters
in their regional temperature system including the temperature anomalies
in the atmosphere of the three regions and the global ocean. \citet{cai2023_JAERE}
also discuss climate impact to economic growth, using projected data
from \citet{burke_large_2018} to calibrate the parameters in measuring
the climate impact. They find that the regional SCC is high in either
a cooperative or a noncooperative world in the presence of climate
damage to economic growth. Moreover, relative to cooperation, noncooperation
reduces the GDP of both economic regions, while the loss in the Tropics
is especially significant. 

An open-loop Nash equilibria might not be as satisfactory as its associated
feedback Nash equilibria in terms of strong time consistency, where
a feedback Nash equilibrium assumes that each agent's strategy depends
on only the current-period state variables. While an open-loop Nash
equilibrium solution could be fairly close to its associated feedback
Nash equilibrium in some cases, it could also be far away in other
cases.\footnote{\citet{Cai_Water_2025} provide examples where an open-loop Nash equilibrium
solution is quite different with its associated feedback Nash equilibrium
solution under a classic lake pollution game.} This feedback Nash equilibrium concept can be associated with the
behavior of countries which enter an international climate agreement
by voluntarily offering to adopt nationally determined emissions paths
as in the Paris Agreement and the Glasgow Climate Pact. \citet{Cai_etal2019_DIRESCU}
design a Dynamic Integrated model of Regional Economy and Spatial
Climate under Uncertainty (DIRESCU), and solve its dynamic stochastic
feedback Nash equilibrium. DIRESCU incorporates the spatial disaggregation
of the economic and climate system of \citet{cai2023_JAERE}, a representative
climate tipping risk and recursive preferences as in DSICE, endogenous
sea level rise, mitigation and adaptation, and permafrost thaw. \citet{Cai_etal2019_DIRESCU}
show that the North has much higher regional carbon taxes than the
Tropics/South. They also find that neglecting heat and moisture transport,
sea level rise, climate tipping risk, or adaptation leads to large
biases in the solutions. 

\subsection{Disaggregation of Sectors}

DICE has only one sector in its economic system. However, except the
spatial heterogeneity, economic sectors face heterogeneous climate
damages, economic growth, abatement costs, etc. Ignoring economic
sector heterogeneity in the economic system will also lead to large
uncertainty in estimation for sector-specific climate policies. \citet{BaldwinCaiKuralbayeva}
extends DICE by disaggregating the global economy into multiple sectors,
including final-goods firms, aggregate-electricity-producing firms,
dirty-electricity-producing firms, fossil-fuel-extracting firms, and
renewable energy firms, where a representative household maximizes
the present value of utilities. The dirty capital stocks used in the
dirty-electricity-producing firms could be underutilized, once they
become uncompetitive, and the ``clean'' sector for the renewable
energy firms is characterized by ``learning-by-doing'': costs of
new technologies decline as a function of cumulative installed capacity
in the clean sector. All firms are assumed to operate under perfect
competition and maximize their profits. The representative household
receives rebates on carbon taxes imposed on the extraction of fossil
fuels, and pays subsidies to renewable energy firms. Except this dynamic
general equilibrium structure, \citet{BaldwinCaiKuralbayeva} also
add one more layer to make their model to be under a principal-agent
framework, where the principal makes decisions on levels of carbon
taxes and subsidies and maximizes the social welfare, bearing in mind
how the other economic participants (the ``agents'' including the
household and the companies) will respond (i.e., subject to the dynamic
general equilibrium conditions). They find that a carbon tax is more
efficient under a stringent climate target that controls the atmospheric
temperature increase in this century smaller than 2$^{\circ}$C, while
a subsidy is more efficient under a mild climate target without the
additional 2$^{\circ}$C restriction. They also find that a portfolio
with both carbon tax and subsidy is the first-best climate policy,
which implements the optimal allocation obtained in the social planner\textquoteright s
problem, while carbon tax only or subsidy only is just a second-best
policy. 

\subsection{Disaggregation of Agents}

DICE or RICE uses global or regional averaging of economic variables,
but climate change impacts are not evenly distributed within the globe
or regions, and poorer people are more vulnerable than the rest of
the population. \citet{dennig_inequality_2015} extend RICE-2010 \citep{Nordhaus_RICE_2010}
to split each of its 12 regions into population quintiles (distributed
by income) to model distributional differences of both consumption
and climate damages within regions. This extended model, called the
Nested Inequalities Climate-Economy (NICE) model, shows that when
future damage falls especially hard on the poor, considerably greater
global mitigation effort is optimal than when damage is proportional
to income. 

DICE posits an infinitely lived social planner to maximize the social
welfare. \citet{Kotlikoff-2021} extend DICE to feature autonomous
overlapping generations and add three dirty energies and one clean
energy in their model. They find that carbon taxation with an appropriate
intergenerational redistribution can make all current and future generations
better off. \citet{KOTLIKOFF2024} further extend the overlapping
generation model of \citet{Kotlikoff-2021} to have multiple regions,
and show that carbon taxation with region and generation-specific
transfers can both correct the carbon externality and raise the welfare
of all current and future agents across all regions.

\section{Overview of Other Key IAMs}

IAMs differ significantly in their structure, assumptions, and applications.
IAMs can be divided into two categories: policy optimization IAMs
and policy evaluation IAMs. A policy optimization IAM is also known
as a cost-benefit IAM, and it includes a damage function mapping temperature
increases to economic damages, allowing the optimal policy to be found
using cost-benefit or cost-effectiveness analysis. A policy evaluation
IAM is also known as a process-based IAM or a simulation IAM, and
it assumes that emissions or mitigation policies are exogenous and
have no feedback to the economy. 

Interagency Working Group on Social Cost of Carbon uses three policy
optimization IAMs---DICE, FUND \citep{Tol1997,AnthoffTol2013}, and
PAGE \citep{hope_page09_2011}---to estimate the SCC. FUND (Climate
Framework for Uncertainty, Negotiation, and Distribution) emphasizes
regional disaggregation with multiple world regions and heterogeneity
in climate impacts. Unlike DICE, FUND models impacts across multiple
sectors and regions, allowing for differential vulnerability. It discusses
probabilistic damage functions and uncertainty in climate sensitivity
and economic parameters. However, per-capita income is assumed to
be exogenous in FUND, while it is endogenous in DICE. PAGE (Policy
Analysis of the Greenhouse Effect) was notably used in the UK government's
Stern Review. PAGE emphasizes probabilistic treatment of uncertainty
by assigning probability distributions to key parameters, such as
damage functions, abatement costs, and climate sensitivity. But FUND
and PAGE are deterministic models and do not discuss noncooperative
equilibria nor risk averse decisions. All the models discussed in
Section 3 are policy optimization models. Other examples of policy
optimization IAMs include WITCH \citep{bosetti_witch_2006}, MERGE
\citep{Manne_merge_2005}, etc. 

Recent policy optimization IAMs disaggregate over space, sectors,
and agents, as shown in the IAMs discussed in Section 3, incorporate
various factors in the economic and climate system, or investigate
various policy instruments \citep{Cai2021}, such as carbon taxation,
cap-and-trade, subsidy, and clean energy standard \citep{Goulder2016}.
For example, \citet{barrage_optimal_2020} builds a dynamic general
equilibrium climate--economy model with distortionary fiscal policy
to quantify optimal carbon taxes. \citet{hambel2021social} extend
RICE to incorporate international trade in a non-cooperative world.
\citet{Fried2022} quantifies the interactions between adaptation,
federal disaster policy, and climate change with a macro heterogeneous-agent
model. \citet{Hong_ECTA2023} investigate adaptation to climate disaster
risks and learning about the disaster arrival frequency, and find
that adaptation is more valuable under learning and that learning
alters SCC projections. \citet{Cruz2024} build a dynamic IAM with
high spatial resolution and discuss the impact of adaptation like
trade, migration, and technological innovations, and the impact of
other policy instruments including carbon taxes, abatement technologies,
and clean energy subsidies. \citet{Kelly2024} find that the optimal
amount of solar geoengineering is very sensitive to belief distributions
about uncertainties of the climate sensitivity and solar geoengineering's
effectiveness. 

\citet{Gillingham_etal_2018} explore uncertainty in baseline trajectories
using six IAMs including three policy optimization IAMs (DICE, FUND,
and WITCH) and three policy evaluation IAMs (GCAM \citep{EDMONDS1983,GCAM_2019},
MERGE \citep{manne_merge_1995,Manne_merge_2005}, and IGSM \citep{chen_long-term_2016}).
They find that parameter uncertainty is more important than model
or structural uncertainty for estimation of key output variables like
the SCC. Other examples of policy evaluation IAMs include IMAGE \citep{IMAGE_2014},
MESSAGE \citep{MESSAGE_2019}, AIM/CGE \citep{AIM_CGE_2017}, REMIND
\citep{REMIND_2015}, etc. These policy evaluation IAMs are more complex,
technology-rich IAMs that incorporate detailed energy system modeling.
These models are frequently used by the Integrated Assessment Modeling
Consortium for scenario development. While not traditionally built
around probabilistic uncertainty, they support ensemble-based scenario
analysis and model comparison to explore uncertainty in pathways.
The SSP and RCP scenarios have enhanced their capacity to capture
socioeconomic uncertainty \citep{RIAHI2017}. However, they are conditional
forecasts and do not assign probabilities, which limits their use
in risk-based decision-making.

\section{Computational Methods}

Computational tractability is always a concern when we build an IAM,
particularly a dynamic stochastic IAM. After we build a complicated
IAM with a lot of efforts, it will be frustrating if it cannot be
solved numerically or accurately. However, computational tractability
depends on which numerical algorithm is applied. 

Policy optimization IAMs are usually non-stationary. To solve a non-stationary
IAM, if it has an infinite time horizon, we often truncate it to be
finite with a terminal time, e.g., 500 years in DICE-2023, and use
a terminal condition or a terminal value function that approximates
the social welfare from the terminal time to infinite. We can choose
a small time horizon for computational tractability if a larger time
horizon has little impact on the results during the period of interest,
which is often the first 100 years or until the end of this century. 

\subsection{Methods for Solving a Deterministic IAM}

If an IAM is a deterministic optimization model, then the most common
method is the so-called optimal control method, which directly applies
a nonlinear programming optimization solver to the truncated finite
horizon IAM, as in DICE. The NEOS server (\citet{NEOS}; \url{https://neos-server.org/neos/solvers/index.html})
provides many efficient optimization solvers. 

For a large-scale IAM, it could be challenging or time-consuming to
solve it using the optimal control method. The starting point strategy
\citep{Cai_etal2012_DICE_CJL} is one efficient way to solve a large-scale
IAM. That is, we solve its corresponding small- or medium-scale IAM
with a larger time step size based on finite difference methods at
first, then use the coarse-time-grid solution and interpolate it over
time to generate a good initial guess for the original large-scale
IAM. Once we have a good initial guess that is close to the true solution,
it is often fast to solve the large optimization problem. 

\subsubsection{Bi-level Optimization}

\citet{BaldwinCaiKuralbayeva} build a principal-agent model, where
the principal decides dynamic carbon taxes and/or subsidies to maximize
the social welfare, and the agents maximize their respective objectives:
the representative household maximizes the present value of household
utilities, and the firms maximize their present values of profits.
For this bi-level optimization problem, \citet{BaldwinCaiKuralbayeva}
use the MPEC (Mathematical Programming with Equilibrium Conditions)
method to transform the bi-level optimization problem to a standard
optimization problem that maximizes the principal's objective subject
to the equilibrium conditions of the agents, and then apply the optimal
control method to solve the transformed optimization problem. MPEC
requires a good initial guess, which can be generated from the social
planner's solution using the optimal control method. For the challenging
problems with the additional investment irreversibility or 2$^{\circ}$C
restriction, \citet{BaldwinCaiKuralbayeva} generate a good initial
guess by applying MPEC to solve their corresponding model without
the additional constraints. 

\subsubsection{Iterative Methods }

It is often challenging to numerically solve large-scale general equilibrium
problems or games with the optimal control method. Particularly for
a dynamic game, those equilibrium conditions might be necessary but
not sufficient for obtaining its true solution. Iterative methods
could be effective to solve them. An iterative method will fix some
variables to some guessed values and transform the large-scale complex
system to some more computationally tractable problems, solve the
transformed problems, then update the guess of the fixed variables
until convergence. 

\citet{cai2023_JAERE} and \citet{CaiMalikShin2023} construct dynamic
games between multiple regions, where each region maximizes their
own regional social welfare, while regional emissions will contribute
to temperature increase and then impact both their own regional output
and other regions'. They solve the corresponding open-loop Nash equilibria
using iterative methods. \citet{cai2023_JAERE} use a social planner's
cooperative solution as an initial guess of regional emission paths.
In the iterations, they solve every region's optimization problem
assuming the other regions' emissions are fixed at the levels at the
previous iteration, then update all regions' emissions with weighted
average of the solution of regional emissions at the current iteration
and the previous until convergence. Similar to \citet{cai2023_JAERE},
the iterative method of \citet{CaiMalikShin2023} updates regional
emissions, amounts of traded emission permits, and permit prices,
until it converges with both the Nash equilibrium of the regions and
the emission trading market's competitive equilibrium. Every region's
optimization problems in the iterations are solved by the optimal
control method. 

\citet{Kotlikoff-2021} and \citet{KOTLIKOFF2024} construct large-scale
overlapping generation models of climate change and the economy, and
solve them with iterative methods too. For example, \citet{KOTLIKOFF2024}
apply an iterative procedure to find the optimal carbon tax path,
updating each period\textquoteright s carbon tax to that period\textquoteright s
SCC in each iteration.

\subsection{Methods for Deep Uncertainty}

Policymakers often have to face deep uncertainty, where a particular
probability distribution cannot be assigned across models, scenarios,
or parameter values. The most common method to deal with problems
involving deep uncertainty is sensitivity analysis, by choosing different
models, scenarios, or values of an uncertain parameter, and checking
if results are qualitatively robust. See, e.g., \citet{Gillingham_etal_2018}
and \citet{Duan_Science2021}, for comparisons across IAMs. When there
are multiple uncertain parameters, a global sensitivity analysis going
through a number of combinations of values of the uncertain parameters
could be more useful than sensitivity analysis only, as it may produce
nontrivial results (see, e.g., \citet{CaiJuddLontzek2017_DSICE} and
\citet{CaiLontzek2019_DSICE}). When it is too time-consuming to run
global sensitivity analysis, we may apply uncertainty quantification
by choosing a small set of nodes (e.g., a sparse grid) for uncertain
parameters and applying an approximation method to estimate solutions
over the whole domain of the uncertain parameters \citep{harenberg_uncertainty_2019}.
However, none of these methods provides a robust and unique solution
for policymakers. 

IAMs with robust decision-making (e.g., the max-min method and the
min-max regret method) help policymakers understand which strategies
are least sensitive to errors in assumptions under deep uncertainty.
These insights are crucial for designing adaptive and flexible climate
policies. The max-min method maximizes the minimal welfare across
the uncertain models, scenarios, or parameter values; that is, it
corresponds to the worst case analysis. Thus, the robust decision
from the max-min method is often too conservative. The min-max regret
method is less conservative. It defines regret to be the difference
between the maximal welfare using the optimal decisions under the
true model and the realized welfare using the proposed decisions under
the other models, then chooses a robust decision to minimize the maximal
regret. For example, \citet{Iverson2012} applies the min-max regret
method to climate policy analysis using DICE-2007 under deep uncertainty
across weights on environmental or growth objectives, climate sensitivity,
and the coefficient of the damage function of DICE. \citet{CaiSanstad2016}
introduce an efficient computational method to solve min-max regret
problems and make robust decisions over deep uncertainty, and apply
it to the Goulder-Mathai model \citep{GoulderMathai2000} for studying
carbon emissions abatement from the energy sector in the face of model
uncertainty about technical change. \citet{Cai_etal2017_FoodPolicy}
apply the computational min-max regret method to obtain the robust
solution of optimal investments in research and development for agricultural
productivity in the face of uncertainty of SSP scenarios. 

An ambiguity-averse individual would rather choose an alternative
with a known probability distribution over one where the probabilities
are unknown. \citet{HansenSargent2008} introduce a robust control
framework with risk and ambiguity aversion, which models utility as
a sum of the current-period utility and the discounted certainty equivalent
of the next-period continuation utility, where the certainty equivalent
is computed using a concave transformation. This robustness approach
has been applied in the literature of IAMs. For example, \citet{Rudik_2020}
incorporates the robust control framework to study the impact of Bayesian
learning on uncertain climate damage. \citet{barnett_brock_hansen_2020}
study risk, ambiguity, and misspecification in continuous-time models
with recursive preferences to assess how alternative uncertainty components
are reflected in valuation of the SCC. Instead of the robustness approach
of \citet{HansenSargent2008}, \citet{Zhao_MS_2023} introduce a full-path
accumulated robustness approach to represent utility as a concave
transformation of the time separable additive von Neumann--Morgenstern
discounted utility, and apply it to a dynamic stochastic IAM with
persistent endogenous discrete disaster states. 

Sometimes knowledge of the exact values of uncertain parameters can
be expressed by some probability distributions, which are referred
to as belief distributions. If uncertain parameters are given with
belief distributions, then Monte Carlo methods are also often used
to generate probabilistic distributions of key output variables in
IAMs, by drawing samples of the uncertain parameters from their belief
distributions, and solving the deterministic model with each sampled
realization of the uncertain parameters. For example, \citet{hope_page09_2011},
\citet{AnthoffTol2013}, and \citet{Gillingham_etal_2018} implement
Monte Carlo methods to analyze the impact of parameter uncertainty
on climate policy. However, Monte Carlo methods do not impose an expectation
operator inside an IAM with parameter uncertainty, that is, it ignores
uncertainty aversion. To incorporate uncertainty aversion, \citet{CaiSanstad2016}
use an expected cost minimization method to find a robust mitigation
pathway in the face of research and development technology uncertainty
with a belief distribution.

If new data can be collected or observed for updating belief distributions
of uncertain parameters, then Bayesian learning can be applied by
shrinking the range of values of the uncertain parameters or reducing
the variances of the belief distributions. Bayesian learning has been
applied in climate change economics, such as \citet{KellyKolstad1999},
\citet{KellyTan2015}, \citet{gerlagh_carbon_2018}, \citet{Rudik_2020},
and \citet{Kelly2024}. Bayesian updating frameworks and ensemble
simulations are used to explore how future information might shift
current policy.

\subsection{Methods for Risk}

\subsubsection{Value Function Iteration}

To deal with risk or Bayesian learning in IAMs, value function iteration
(VFI) is the most common method to solve a dynamic stochastic IAM.
Using DSICE as an example, \citet{CaiJuddLontzek2017_DSICE} and \citet{CaiLontzek2019_DSICE}
apply VFI to solve the following Bellman equation: 
\begin{eqnarray*}
V_{t}(\mathbf{S})=\max_{C,\mu} &  & u(C,L_{t})+\frac{\beta}{\Xi}\left[\mathbb{E}_{t}\left\{ \left(\Xi V_{t+1}\left(\mathbf{S}^{+}\right)\right)^{\varTheta}\right\} \right]^{\frac{1}{\varTheta}},\\
\text{s.t.} &  & K^{+}=(1-\delta)K+\widehat{Y}_{t}-C,\\
 &  & \mathbf{M}^{+}=\Phi_{M}\mathbf{M}+\left(E_{t},0,0\right)^{\top},\\
 &  & \mathbf{T}^{+}=\Phi_{T}\mathbf{T}+\left(\xi_{1}\mathcal{F}_{t}\left(M_{\mathrm{AT}}\right),0\right)^{\top},\\
 &  & \zeta^{+}=g_{\zeta}(\zeta,\chi,\omega_{\zeta}),\\
 &  & \chi^{+}=g_{\chi}(\chi,\omega_{\chi}),\\
 &  & J^{+}=g_{J}(J,\mathbf{T},\omega_{J})
\end{eqnarray*}
where $V_{t}$ is the value function at time $t$, $\mathbf{S}=(K,\mathbf{M},\mathbf{T},\zeta,\chi,J)$
is a nine-dimensional state vector, $\mathbf{S}^{+}=(K^{+},\mathbf{M}^{+},\mathbf{T}^{+},\zeta^{+},\chi^{+},J^{+})$
is the next state vector, and the transition laws $\zeta^{+}=g_{\zeta}(\zeta,\chi,\omega_{\zeta})$
and $\chi^{+}=g_{\chi}(\chi,\omega_{\chi})$ represent the dense Markov
chains discretized from the long-run growth risk (\ref{eq:zeta_process})-(\ref{eq:chi_process}).
The original infinite horizon is truncated to 600 years as in DICE-2007,
and the terminal value function $V_{601}$ is constructed as 
\begin{equation}
V_{601}(\mathbf{S})=\sum_{t=601}^{1000}\beta^{t-601}u(C_{t},L_{t})\label{eq:terminalV}
\end{equation}
where $C_{t}=0.78\widehat{Y}_{t}$ for $601\leq t\leq1000$, assuming
\textcolor{black}{that after the terminal time the system is deterministic
and stationary with zero emissions. }

With the terminal value function $V_{601}$, VFI iterates backward
over time to get all value functions and policy functions, which require
numerical approximation over the state space. DSICE uses complete
Chebyshev polynomial approximation: the values of $V_{t}$ at tensor
Chebyshev nodes on time-varying approximation domains on the state
space are computed via numerical optimization in the Bellman equation
in parallel \citep{Caietal2015_parallel}, and these values are used
for calculating coefficients of the complete Chebyshev polynomials
(see \citet{Cai2019} for a more detailed discussion). 

DSICE is a non-stationary dynamic stochastic model. If it is transformed
into a stationary model by adding additional states including those
time-varying exogenous parameters or time, like what \citet{LemoineTraeger2014}
did, it will have a higher-dimension state space and a much wider
approximation domain on the state space, then it will require much
higher degree Chebyshev polynomials for numerical approximation, implying
many more approximation nodes and their associated optimization problems.
Moreover, even with a high degree Chebyshev polynomial, it could still
be challenging to achieve a high-accuracy approximation, as it would
often have to impose an additional restriction that next states are
not beyond the approximation domain. For example, if we add time as
an additional state variable, then next state of time will exceed
the upper bound of the time state variable if the current time state
is at the upper bound. The additional restriction creates additional
kinks for the value function approximation, reduces accuracy of approximation,
and even makes it challenging to numerically solve the optimization
problems in the Bellman equation. Therefore, it will be much more
efficient and accurate if we use time-varying approximation domains
to solve the non-stationary model directly, without transforming it
to be stationary. For example, after the use of time-varying approximation
domains, the tipping benchmark examples of DSICE require just degree-four
complete Chebyshev polynomials at each time in VFI to achieve a high-accuracy
solution. To construct the time-varying approximation domains, we
can start \textcolor{black}{with a narrow domain at the initial time
and iteratively choose the time $t+1$ domain so that any combination
of time $t$ states, time $t$ optimal action, and time $t$ shocks
will be transited to a point inside the time $t+1$ domain according
to the transition laws. }

It is also critical to verify whether VFI solves a dynamic stochastic
model accurately. One way is to apply the same code of VFI to solve
a nearly deterministic version of the dynamic stochastic model by
reducing all randomness to have nearly zero variances, and then verify
whether its solution is close to the solution of the deterministic
version, which can be obtained by the optimal control method. In addition,
it is also necessary to check whether numerical approximation errors
are small or whether a higher degree approximation with more approximation
nodes has little impact on the solution. 

\subsubsection{Time Iteration}

It is also common to use time iteration, also known as policy function
iteration, for solving dynamic stochastic problems. Time iteration
constructs policy functions of state variables on the approximation
domains of the state space, by solving a system of constraints including
intertemporal Euler equations and transition laws, and other first-order
or Karush-Kuhn-Tucker conditions. However, since the system of constraints
are necessary but not sufficient conditions for the original dynamic
optimization problem, we should always check whether the converged
solution from time iteration is unique. Moreover, time iteration cannot
solve problems when some decision variables are discrete or when the
first derivatives over continuous decision variables do not exist
at some points of objective or constraint functions. In the literature,
time iteration has been applied to solve IAMs, particularly dynamic
stochastic general equilibrium or dynamic stochastic game problems.
For example, \citet{Cai_etal2019_DIRESCU} combine VFI and time iteration
to solve their feedback Nash equilibrium problems with recursive preferences. 

\subsubsection{NLCEQ}

It is often challenging to apply VFI or time iteration to solve dynamic
stochastic IAMs with high dimensions or occasionally binding constraints.
\citet{Cai_etal_QE2017} introduce a Non-Linear Certainty Equivalent
approximation method (NLCEQ) to solve these kinds of problems with
acceptable accuracy, including a social planner's problems and competitive
equilibrium. The algorithm is simple for coding, naturally parallelizable,
and is also very stable, particularly for solving a social planner's
problems like many IAMs. For example, NLCEQ can be applied to solve
variants of DSICE with time-separable utility. Moreover, NLCEQ can
generate a policy function for stationary problems or policy functions
at any periods of interest for non-stationary problems. Therefore,
even if NLCEQ might not be able to provide accurate solutions for
some IAMs like DSICE with recursive preferences, we may use NLCEQ
to generate a terminal value or policy function, and then apply VFI
or time iteration to iterate backward over time, while the terminal
time does not have to be much larger than the period of interest. 

\subsubsection{SCEQ}

After VFI, time iteration, or NLCEQ solves value and policy functions
at every period, we often need to use the value and policy functions
to do a forward simulation process for generating distributions of
future key output variables for policymaking. However, it is often
challenging to have an accurate numerical approximation to the value
and policy functions when the dimension of the state space is high
or the functions have kinks. \citet{Cai_SCEQ} introduce a Simulated
Certainty Equivalent approximation method (SCEQ) to solve dynamic
stochastic problems by directly generating distributions of future
key output variables without constructing value and policy functions.
They show that SCEQ can quickly solve high-dimensional dynamic stochastic
problems with hundreds of state variables, a wide state space, and
occasionally binding constraints, using just a desktop computer. They
also show that SCEQ can efficiently solve two simpler versions of
DSICE, assuming a simple economic risk or climate tipping risk without
using Epstein-Zin preferences. This simple and stable SCEQ algorithm
has been applied to solve a large-scale dynamic stochastic global
land resource use problem with stochastic crop yields due to adverse
climate impacts and limits on further technological progress \citep{Steinbuks-2024}. 

\section{Summary Points and Future Issues}

IAMs are indispensable for understanding the economic implications
of climate change and guiding climate policy. However, their usefulness
hinges on how they handle uncertainty. As this review shows, uncertainty
arises at multiple levels---from parameter values to structural and
deep uncertainties and risks---and has profound effects on policy
recommendations. Recent papers suggest significant advances are needed
to deal with complex uncertainty structures, especially in light of
increasing extreme weathers, new technological pathways, and global
political shifts. Advances in stochastic modeling, robust optimization,
and equity considerations enhance their relevance for policymaking.
Recent methodological advances offer powerful tools for integrating
uncertainty into IAMs, and improve our ability to design climate strategies
that are both risk-informed, dynamically adaptive, and robust. 

Future work should continue expanding the scope of uncertainty represented
in IAMs (e.g., finer disaggregation of space, sectors, and agents),
improving the computational tools (such as deep learning methods reviewed
in \citet{Fernandez-Villaverde2025}) for solving IAMs, and analyzing
various climate policies. Moreover, future IAMs could incorporate
richer behavioral and heterogeneity representations, demand-side mitigation
\citep{Creutzig2022}, human capital \citep{Paudel2025}, ecosystems
\citep{Johnson2025}, carbon sequestration \citep{Brent-2020,golub_ERL_2022},
carbon dioxide removal \citep{beerling_Nature_2020,Edenhofer2025},
and artificial intelligence technologies \citep{Khanna-2024}. Furthermore,
future IAMs would consider planetary boundaries \citep{Hertel2025},
evaluation of sustainability, equity, and resilience \citep{Liu_science_2015,irwin_welfare_2016,Baylis-2021,Koundouri2025,Tibebu_2025},
and the nexus of food, energy, and water systems \citep{Kling-2017,Miao-2020},
in the face of uncertainty and climate change. Finally, future work
may also include synergies between policy optimization IAMs and policy
evaluation IAMs \citep{Fisher-Vanden-2020}.

\bibliographystyle{jpe}
\bibliography{Cai_ARRE_2025}

\begin{thebibliography}{116}
\newcommand{\enquote}[1]{``#1''}
\providecommand{\natexlab}[1]{#1}
\providecommand{\url}[1]{\texttt{#1}}
\providecommand{\urlprefix}{URL }

\bibitem[{Anthoff and Tol(2013)}]{AnthoffTol2013}
Anthoff, David and Richard Tol. 2013.
\newblock \enquote{The uncertainty about the social cost of carbon: A
  decomposition analysis using {FUND}.}
\newblock \emph{Climatic Change} 117~(3):515--530.

\bibitem[{Baldwin, Cai, and Kuralbayeva(2020)}]{BaldwinCaiKuralbayeva}
Baldwin, Elizabeth, Yongyang Cai, and Karlygash Kuralbayeva. 2020.
\newblock \enquote{To Build or Not to Build? Capital Stocks and Climate
  Policy.}
\newblock \emph{Journal of Environmental Economics and Management} 100~(Article
  102235).

\bibitem[{Barnett, Brock, and Hansen(2020)}]{barnett_brock_hansen_2020}
Barnett, Michael, William~A. Brock, and Lars~Peter Hansen. 2020.
\newblock \enquote{Pricing {Uncertainty} {Induced} by {Climate} {Change}.}
\newblock \emph{The Review of Financial Studies} 33~(3):1024--1066.

\bibitem[{Barrage(2020)}]{barrage_optimal_2020}
Barrage, Lint. 2020.
\newblock \enquote{Optimal {Dynamic} {Carbon} {Taxes} in a {Climate-Economy}
  {Model} with {Distortionary} {Fiscal} {Policy}.}
\newblock \emph{The Review of Economic Studies} 87~(1):1--39.

\bibitem[{Barrage and Nordhaus(2024)}]{BarrageNordhaus2023}
Barrage, Lint and William Nordhaus. 2024.
\newblock \enquote{Policies, projections, and the social cost of carbon:
  Results from the DICE-2023 model.}
\newblock \emph{Proceedings of the National Academy of Sciences}
  121~(13):e2312030121.

\bibitem[{Baylis, Heckelei, and Hertel(2021)}]{Baylis-2021}
Baylis, Kathy, Thomas Heckelei, and Thomas~W. Hertel. 2021.
\newblock \enquote{Agricultural Trade and Environmental Sustainability.}
\newblock \emph{Annual Review of Resource Economics} 13:379--401.

\bibitem[{Beerling et~al.(2020)Beerling, Kantzas, Lomas, Wade, Eufrasio,
  Renforth, Sarkar, Andrews, James, Pearce, Mercure, Pollitt, Holden, Edwards,
  Khanna, Koh, Quegan, Pidgeon, Janssens, Hansen, and
  Banwart}]{beerling_Nature_2020}
Beerling, David~J., Euripides~P. Kantzas, Mark~R. Lomas, Peter Wade, Rafael~M.
  Eufrasio, Phil Renforth, Binoy Sarkar, M.~Grace Andrews, Rachael~H. James,
  Christopher~R. Pearce, Jean-Francois Mercure, Hector Pollitt, Philip~B.
  Holden, Neil~R. Edwards, Madhu Khanna, Lenny Koh, Shaun Quegan, Nick~F.
  Pidgeon, Ivan~A. Janssens, James Hansen, and Steven~A. Banwart. 2020.
\newblock \enquote{Potential for large-scale {CO2} removal via enhanced rock
  weathering with croplands.}
\newblock \emph{Nature} 583~(7815):242--248.

\bibitem[{Bilal and Stock(2025)}]{Bilal2025}
Bilal, Adrien and James~H. Stock. 2025.
\newblock \enquote{A Guide to Macroeconomics and Climate Change.}
\newblock NBER Working Paper 33567.
\newblock \urlprefix\url{https://www.nber.org/papers/w33567}.

\bibitem[{Bohringer et~al.(2022)Bohringer, Fischer, Rosendahl, and
  Rutherford}]{bohringer_NCC_2022}
Bohringer, Christoph, Carolyn Fischer, Knut~Einar Rosendahl, and Thomas~Fox
  Rutherford. 2022.
\newblock \enquote{Potential impacts and challenges of border carbon
  adjustments.}
\newblock \emph{Nature Climate Change} 12~(1):22--29.

\bibitem[{Bosetti(2021)}]{Bosetti2021}
Bosetti, Valentina. 2021.
\newblock \enquote{Integrated Assessment Models for Climate Change.}
\newblock \emph{Oxford Research Encyclopedia of Economics and Finance}
  \urlprefix\url{https://doi.org/10.1093/acrefore/9780190625979.013.572}.

\bibitem[{Bosetti et~al.(2006)Bosetti, Carraro, Galeotti, Massetti, and
  Tavoni}]{bosetti_witch_2006}
Bosetti, Valentina, Carlo Carraro, Marzio Galeotti, Emanuele Massetti, and
  Massimo Tavoni. 2006.
\newblock \enquote{{WITCH} {A} {World} {Induced} {Technical} {Change} {Hybrid}
  {Model}.}
\newblock \emph{The Energy Journal} 27:13--37.

\bibitem[{Burke, Davis, and Diffenbaugh(2018)}]{burke_large_2018}
Burke, Marshall, W.~Matthew Davis, and Noah~S. Diffenbaugh. 2018.
\newblock \enquote{Large potential reduction in economic damages under {UN}
  mitigation targets.}
\newblock \emph{Nature} 557~(7706):549.

\bibitem[{Cai(2019)}]{Cai2019}
Cai, Yongyang. 2019.
\newblock \enquote{Computational methods in environmental and resource
  economics.}
\newblock \emph{Annual Review of Resource Economics} 11:59--82.

\bibitem[{Cai(2021)}]{Cai2021}
---{}---{}---. 2021.
\newblock \enquote{The Role of Uncertainty in Controlling Climate Change.}
\newblock \emph{Oxford Research Encyclopedia of Economics and Finance}
  \urlprefix\url{https://doi.org/10.1093/acrefore/9780190625979.013.573}.

\bibitem[{Cai, Brock, and Xepapadeas(2023)}]{cai2023_JAERE}
Cai, Yongyang, William Brock, and Anastasios Xepapadeas. 2023.
\newblock \enquote{Climate change impact on economic growth: Regional climate
  policy under cooperation and noncooperation.}
\newblock \emph{Journal of the Association of Environmental and Resource
  Economists} 10~(3):569--605.

\bibitem[{Cai et~al.(2019)Cai, Brock, Xepapadeas, and
  Judd}]{Cai_etal2019_DIRESCU}
Cai, Yongyang, William Brock, Anastasios Xepapadeas, and Kenneth~L. Judd. 2019.
\newblock \enquote{Climate Policy under Spatial Heat Transport: Cooperative and
  Noncooperative Regional Outcomes.}
\newblock arXiv Working Paper 1909.04009.
\newblock \urlprefix\url{https://doi.org/10.48550/arXiv.1909.04009}.

\bibitem[{Cai, Golub, and Hertel(2017)}]{Cai_etal2017_FoodPolicy}
Cai, Yongyang, Alla~A. Golub, and Thomas~W. Hertel. 2017.
\newblock \enquote{Agricultural research spending must increase in light of
  future uncertainties.}
\newblock \emph{Food Policy} 70:71--83.

\bibitem[{Cai and Judd(2023)}]{Cai_SCEQ}
Cai, Yongyang and Kenneth~L. Judd. 2023.
\newblock \enquote{A simple but powerful simulated certainty equivalent
  approximation method for dynamic stochastic problems.}
\newblock \emph{Quantitative Economics} 14~(2):651--687.

\bibitem[{Cai et~al.(2015{\natexlab{a}})Cai, Judd, Lenton, Lontzek, and
  Narita}]{Cai_PNAS2015}
Cai, Yongyang, Kenneth~L. Judd, Timothy~M. Lenton, Thomas~S. Lontzek, and Daiju
  Narita. 2015{\natexlab{a}}.
\newblock \enquote{Environmental tipping points significantly affect the
  cost-benefit assessment of climate policies.}
\newblock \emph{Proceedings of the National Academy of Sciences}
  112~(15):4606--4611.

\bibitem[{Cai, Judd, and Lontzek(2012{\natexlab{a}})}]{Cai_etal2012_DICE_CJL}
Cai, Yongyang, Kenneth~L. Judd, and Thomas~S. Lontzek. 2012{\natexlab{a}}.
\newblock \enquote{Continuous-time methods for integrated assessment models.}
\newblock NBER Working Paper 18365.
\newblock \urlprefix\url{https://www.nber.org/papers/w18365}.

\bibitem[{Cai, Judd, and Lontzek(2012{\natexlab{b}})}]{cai_open_2012}
---{}---{}---. 2012{\natexlab{b}}.
\newblock \enquote{Open science is necessary.}
\newblock \emph{Nature Climate Change} 2~(5):299--299.

\bibitem[{Cai, Judd, and Lontzek(2017)}]{CaiJuddLontzek2017_DSICE}
---{}---{}---. 2017.
\newblock \enquote{The social cost of carbon with economic and climate risks.}
\newblock Hoover economics working paper 18113.
\newblock
  \urlprefix\url{https://www.hoover.org/research/social-cost-carbon-economic-and-climate-risk}.

\bibitem[{Cai, Judd, and Steinbuks(2017)}]{Cai_etal_QE2017}
Cai, Yongyang, Kenneth~L. Judd, and Jevgenijs Steinbuks. 2017.
\newblock \enquote{A nonlinear certainty equivalent approximation method for
  dynamic stochastic problems.}
\newblock \emph{Quantitative Economics} 8~(1):117--147.

\bibitem[{Cai et~al.(2015{\natexlab{b}})Cai, Judd, Thain, and
  Wright}]{Caietal2015_parallel}
Cai, Yongyang, Kenneth~L. Judd, Greg Thain, and Steven Wright.
  2015{\natexlab{b}}.
\newblock \enquote{Solving dynamic programming problems on computational grid.}
\newblock \emph{Computational Economics} 45~(2):261--284.

\bibitem[{Cai, Lenton, and Lontzek(2016)}]{cai_NCC_2016}
Cai, Yongyang, Timothy~M. Lenton, and Thomas~S. Lontzek. 2016.
\newblock \enquote{Risk of multiple interacting tipping points should encourage
  rapid {CO}2 emission reduction.}
\newblock \emph{Nature Climate Change} 6~(5):520--525.

\bibitem[{Cai and Lontzek(2019)}]{CaiLontzek2019_DSICE}
Cai, Yongyang and Thomas~S. Lontzek. 2019.
\newblock \enquote{The social cost of carbon with economic and climate risks.}
\newblock \emph{Journal of Political Economy} 6:2684--2734.

\bibitem[{Cai, Malik, and Shin(2023)}]{CaiMalikShin2023}
Cai, Yongyang, Khyati Malik, and Hyeseon Shin. 2023.
\newblock \enquote{Dynamics of Global Emission Permit Prices and Regional
  Social Cost of Carbon under Noncooperation.}
\newblock arXiv Working Paper 2312.15563.
\newblock \urlprefix\url{https://doi.org/10.48550/arXiv.2312.15563}.

\bibitem[{Cai and Sanstad(2016)}]{CaiSanstad2016}
Cai, Yongyang and Alan~H. Sanstad. 2016.
\newblock \enquote{Model uncertainty and energy technology policy: The example
  of induced technical change.}
\newblock \emph{Computers \& Operations Research} 66:362--373.

\bibitem[{Cai, Xepapadeas, and de~Zeeuw(2025)}]{Cai_Water_2025}
Cai, Yongyang, Anastasios Xepapadeas, and Aart de~Zeeuw. 2025.
\newblock \enquote{Solving Nash Equilibria in Nonlinear Differential Games for
  Common-Pool Resources.}
\newblock arXiv Working Paper 2506.06646.
\newblock \urlprefix\url{https://doi.org/10.48550/arXiv.2506.06646}.

\bibitem[{Calvin et~al.(2019)Calvin, Patel, Clarke, Asrar, Bond-Lamberty, Cui,
  Di~Vittorio, Dorheim, Edmonds, Hartin, Hejazi, Horowitz, Iyer, Kyle, Kim,
  Link, McJeon, Smith, Snyder, Waldhoff, and Wise}]{GCAM_2019}
Calvin, Katherine, Pralit Patel, Leon Clarke, Ghassem Asrar, Ben Bond-Lamberty,
  Ryna~Yiyun Cui, Alan Di~Vittorio, Kalyn Dorheim, Jae Edmonds, Corinne Hartin,
  Mohamad Hejazi, Russell Horowitz, Gokul Iyer, Page Kyle, Sonny Kim, Robert
  Link, Haewon McJeon, Steven~J. Smith, Abigail Snyder, Stephanie Waldhoff, and
  Marshall Wise. 2019.
\newblock \enquote{{GCAM} v5.1: representing the linkages between energy,
  water, land, climate, and economic systems.}
\newblock \emph{Geoscientific Model Development} 12~(2):677--698.

\bibitem[{Chen et~al.(2016)Chen, Paltsev, Reilly, Morris, and
  Babiker}]{chen_long-term_2016}
Chen, Y. H.~Henry, Sergey Paltsev, John~M. Reilly, Jennifer~F. Morris, and
  Mustafa~H. Babiker. 2016.
\newblock \enquote{Long-term economic modeling for climate change assessment.}
\newblock \emph{Economic Modelling} 52:867--883.

\bibitem[{Creutzig et~al.(2022)Creutzig, Niamir, Bai, Callaghan, Cullen,
  Diaz-Jose, Figueroa, Grubler, Lamb, Leip, Masanet, Mata, Mattauch, Minx,
  Mirasgedis, Mulugetta, Nugroho, Pathak, Perkins, Roy, de~la Rue~du Can,
  Saheb, Some, Steg, Steinberger, and Urge-Vorsatz}]{Creutzig2022}
Creutzig, Felix, Leila Niamir, Xuemei Bai, Max Callaghan, Jonathan Cullen,
  Julio Diaz-Jose, Maria Figueroa, Arnulf Grubler, William~F. Lamb, Adrian
  Leip, Eric Masanet, Erika Mata, Linus Mattauch, Jan~C. Minx, Sebastian
  Mirasgedis, Yacob Mulugetta, Sudarmanto~Budi Nugroho, Minal Pathak, Patricia
  Perkins, Joyashree Roy, Stephane de~la Rue~du Can, Yamina Saheb, Shreya Some,
  Linda Steg, Julia Steinberger, and Diana Urge-Vorsatz. 2022.
\newblock \enquote{Demand-side solutions to climate change mitigation
  consistent with high levels of well-being.}
\newblock \emph{Nature Climate Change} 12:36--46.

\bibitem[{Cruz and Rossi-Hansberg(2024)}]{Cruz2024}
Cruz, Jose-Luis and Esteban Rossi-Hansberg. 2024.
\newblock \enquote{The Economic Geography of Global Warming.}
\newblock \emph{The Review of Economic Studies} 91~(2):899--939.

\bibitem[{Czyzyk, Mesnier, and More(1998)}]{NEOS}
Czyzyk, J., M.~P. Mesnier, and J.~J. More. 1998.
\newblock \enquote{The NEOS server.}
\newblock \emph{IEEE Computational Science \& Engineering} 5:68--75.

\bibitem[{Dennig et~al.(2015)Dennig, Budolfson, Fleurbaey, Siebert, and
  Socolow}]{dennig_inequality_2015}
Dennig, Francis, Mark~B. Budolfson, Marc Fleurbaey, Asher Siebert, and
  Robert~H. Socolow. 2015.
\newblock \enquote{Inequality, climate impacts on the future poor, and carbon
  prices.}
\newblock \emph{Proceedings of the National Academy of Sciences}
  112~(52):15827--15832.

\bibitem[{Desmet and Rossi-Hansberg(2024)}]{Desmet2024}
Desmet, Klaus and Esteban Rossi-Hansberg. 2024.
\newblock \enquote{Climate Change Economics over Time and Space.}
\newblock \emph{Annual Review of Economics} 16:271--304.

\bibitem[{Dietz(2024)}]{DIETZ2024}
Dietz, Simon. 2024.
\newblock \enquote{Chapter 1 - Introduction to integrated assessment modeling
  of climate change.}
\newblock In \emph{Handbook of the Economics of Climate Change}, vol.~1, edited
  by Lint Barrage and Solomon Hsiang. North-Holland, 1--51.

\bibitem[{Dietz et~al.(2021)Dietz, Rising, Stoerk, and
  Wagner}]{Dietz_PNAS_2021}
Dietz, Simon, James Rising, Thomas Stoerk, and Gernot Wagner. 2021.
\newblock \enquote{Economic impacts of tipping points in the climate system.}
\newblock \emph{Proceedings of the National Academy of Sciences}
  118~(34):e2103081118.

\bibitem[{Dietz and Stern(2015)}]{DietzStern2015}
Dietz, Simon and Nicholas Stern. 2015.
\newblock \enquote{Endogenous Growth, Convexity of Damage and Climate Risk: How
  Nordhaus' Framework Supports Deep Cuts in Carbon Emissions.}
\newblock \emph{The Economic Journal} 125~(583):574--620.

\bibitem[{Drupp et~al.(2018)Drupp, Freeman, Groom, and
  Nesje}]{drupp_discounting_2018}
Drupp, Moritz~A., Mark~C. Freeman, Ben Groom, and Frikk Nesje. 2018.
\newblock \enquote{Discounting {Disentangled}.}
\newblock \emph{American Economic Journal: Economic Policy} 10~(4):109--134.

\bibitem[{Duan et~al.(2021)Duan, Zhou, Jiang, Bertram, Harmsen, Kriegler, van
  Vuuren, Wang, Fujimori, Tavoni, Ming, Keramidas, Iyer, and
  Edmonds}]{Duan_Science2021}
Duan, Hongbo, Sheng Zhou, Kejun Jiang, Christoph Bertram, Mathijs Harmsen,
  Elmar Kriegler, Detlef~P. van Vuuren, Shouyang Wang, Shinichiro Fujimori,
  Massimo Tavoni, Xi~Ming, Kimon Keramidas, Gokul Iyer, and James Edmonds.
  2021.
\newblock \enquote{Assessing China's efforts to pursue the 1.5C warming limit.}
\newblock \emph{Science} 372~(6540):378--385.

\bibitem[{Edenhofer et~al.(2025)Edenhofer, Franks, Gruner, Kalkuhl, and
  Lessmann}]{Edenhofer2025}
Edenhofer, Ottmar, Max Franks, Friedemann Gruner, Matthias Kalkuhl, and Kai
  Lessmann. 2025.
\newblock \enquote{The Economics of Carbon Dioxide Removal.}
\newblock \emph{Annual Review of Resource Economics} 17:301--321.

\bibitem[{Edmonds and Reilly(1983)}]{EDMONDS1983}
Edmonds, Jae and John Reilly. 1983.
\newblock \enquote{A long-term global energy- economic model of carbon dioxide
  release from fossil fuel use.}
\newblock \emph{Energy Economics} 5~(2):74--88.

\bibitem[{Epstein and Zin(1989)}]{Epstein-Zin-1989}
Epstein, Larry~G. and Stanley~E. Zin. 1989.
\newblock \enquote{Substitution, Risk Aversion, and the Temporal Behavior of
  Consumption and Asset Returns: A Theoretical Framework.}
\newblock \emph{Econometrica} 57~(4):937--969.

\bibitem[{Fernandez-Villaverde, Gillingham, and
  Scheidegger(2025)}]{Fernandez-Villaverde2025}
Fernandez-Villaverde, Jesus, Kenneth~T. Gillingham, and Simon Scheidegger.
  2025.
\newblock \enquote{Climate Change Through the Lens of Macroeconomic Modeling.}
\newblock \emph{Annual Review of Economics} 17:125--150.

\bibitem[{Fisher-Vanden and Weyant(2020)}]{Fisher-Vanden-2020}
Fisher-Vanden, Karen and John Weyant. 2020.
\newblock \enquote{The Evolution of Integrated Assessment: Developing the Next
  Generation of Use-Inspired Integrated Assessment Tools.}
\newblock \emph{Annual Review of Resource Economics} 12:471--487.

\bibitem[{Folini et~al.(2025)Folini, Friedl, Kubler, and
  Scheidegger}]{Folini2025}
Folini, Doris, Aleksandra Friedl, Felix Kubler, and Simon Scheidegger. 2025.
\newblock \enquote{The Climate in Climate Economics.}
\newblock \emph{The Review of Economic Studies} 92~(1):299--338.

\bibitem[{Fowlie and Reguant(2022)}]{fowlie2022mitigating}
Fowlie, Meredith~L and Mar Reguant. 2022.
\newblock \enquote{Mitigating emissions leakage in incomplete carbon markets.}
\newblock \emph{Journal of the Association of Environmental and Resource
  Economists} 9~(2):307--343.

\bibitem[{Fried(2022)}]{Fried2022}
Fried, Stephie. 2022.
\newblock \enquote{Seawalls and Stilts: A Quantitative Macro Study of Climate
  Adaptation.}
\newblock \emph{The Review of Economic Studies} 89~(6):3303--3344.

\bibitem[{Fujimori, Masui, and Matsuoka(2017)}]{AIM_CGE_2017}
Fujimori, Shinichiro, Toshihiko Masui, and Yuzuru Matsuoka. 2017.
\newblock \enquote{{AIM}/{CGE} {V2}.0 {Model} {Formula}.}
\newblock In \emph{Post-2020 {Climate} {Action}: {Global} and {Asian}
  {Perspectives}}, edited by Shinichiro Fujimori, Mikiko Kainuma, and Toshihiko
  Masui. Singapore: Springer, 201--303.

\bibitem[{Gerlagh and Liski(2018)}]{gerlagh_carbon_2018}
Gerlagh, Reyer and Matti Liski. 2018.
\newblock \enquote{Carbon {Prices} for {The} {Next} {Hundred} {Years}.}
\newblock \emph{The Economic Journal} 128~(609):728--757.

\bibitem[{Gillingham et~al.(2018)Gillingham, Nordhaus, Anthoff, Blanford,
  Bosetti, Christensen, McJeon, and Reilly}]{Gillingham_etal_2018}
Gillingham, Kenneth, William Nordhaus, David Anthoff, Geoffrey Blanford,
  Valentina Bosetti, Peter Christensen, Haewon McJeon, and John Reilly. 2018.
\newblock \enquote{Modeling Uncertainty in Integrated Assessment of Climate
  Change: A Multimodel Comparison.}
\newblock \emph{Journal of the Association of Environmental and Resource
  Economists} 5~(4):791--826.

\bibitem[{Golosov et~al.(2014)Golosov, Hassler, Krusell, and
  Tsyvinski}]{Golosov2014}
Golosov, Mikhail, John Hassler, Per Krusell, and Aleh Tsyvinski. 2014.
\newblock \enquote{Optimal Taxes on Fossil Fuel in General Equilibrium.}
\newblock \emph{Econometrica} 82~(1):41--88.

\bibitem[{Golub et~al.(2022)Golub, Sohngen, Cai, Kim, and
  Hertel}]{golub_ERL_2022}
Golub, Alla, Brent Sohngen, Yongyang Cai, John Kim, and Thomas Hertel. 2022.
\newblock \enquote{Costs of forest carbon sequestration in the presence of
  climate change impacts.}
\newblock \emph{Environmental Research Letters} 17~(10):104011.
\newblock Publisher: IOP Publishing.

\bibitem[{Goulder, Hafstead, and Williams(2016)}]{Goulder2016}
Goulder, Lawrence~H., Marc A.~C. Hafstead, and Roberton~C. Williams. 2016.
\newblock \enquote{General Equilibrium Impacts of a Federal Clean Energy
  Standard.}
\newblock \emph{American Economic Journal: Economic Policy} 8~(2):186--218.

\bibitem[{Goulder et~al.(2022)Goulder, Long, Lu, and
  Morgenstern}]{goulder2022china}
Goulder, Lawrence~H, Xianling Long, Jieyi Lu, and Richard~D Morgenstern. 2022.
\newblock \enquote{China's unconventional nationwide CO2 emissions trading
  system: Cost-effectiveness and distributional impacts.}
\newblock \emph{Journal of Environmental Economics and Management} 111:102561.

\bibitem[{Goulder and Mathai(2000)}]{GoulderMathai2000}
Goulder, Lawrence~H. and Koshy Mathai. 2000.
\newblock \enquote{Optimal CO2 Abatement in the Presence of Induced
  Technological Change.}
\newblock \emph{Journal of Environmental Economics and Management}
  39~(1):1--38.

\bibitem[{Hambel, Kraft, and Schwartz(2021{\natexlab{a}})}]{HAMBEL_EER2021}
Hambel, Christoph, Holger Kraft, and Eduardo Schwartz. 2021{\natexlab{a}}.
\newblock \enquote{Optimal carbon abatement in a stochastic equilibrium model
  with climate change.}
\newblock \emph{European Economic Review} 132:103642.

\bibitem[{Hambel, Kraft, and Schwartz(2021{\natexlab{b}})}]{hambel2021social}
---{}---{}---. 2021{\natexlab{b}}.
\newblock \enquote{The social cost of carbon in a non-cooperative world.}
\newblock \emph{Journal of International Economics} 131:103490.

\bibitem[{Hansen and Sargent(2008)}]{HansenSargent2008}
Hansen, Lars~Peter and Thomas Sargent. 2008.
\newblock \emph{Robustness}.
\newblock Princeton, NJ: Princeton University Press.

\bibitem[{Harenberg et~al.(2019)Harenberg, Marelli, Sudret, and
  Winschel}]{harenberg_uncertainty_2019}
Harenberg, Daniel, Stefano Marelli, Bruno Sudret, and Viktor Winschel. 2019.
\newblock \enquote{Uncertainty quantification and global sensitivity analysis
  for economic models.}
\newblock \emph{Quantitative Economics} 10~(1):1--41.

\bibitem[{Hertel(2025)}]{Hertel2025}
Hertel, Thomas~W. 2025.
\newblock \enquote{Economic Analysis of Global and Local Policies for
  Respecting Planetary Boundaries.}
\newblock \emph{Agricultural Economics} 56~(3):336--348.

\bibitem[{Hong, Wang, and Yang(2023)}]{Hong_ECTA2023}
Hong, Harrison, Neng Wang, and Jinqiang Yang. 2023.
\newblock \enquote{Mitigating Disaster Risks in the Age of Climate Change.}
\newblock \emph{Econometrica} 91~(5):1763--1802.

\bibitem[{Hope(2011)}]{hope_page09_2011}
Hope, Chris. 2011.
\newblock \enquote{The {PAGE09} integrated assessment model: {A} technical
  description.}
\newblock Working {Paper} 4/2011, Cambridge Judge Business School.
\newblock
  \urlprefix\url{https://www.jbs.cam.ac.uk/fileadmin/user_upload/research/workingpapers/wp1104.pdf}.

\bibitem[{Huppmann et~al.(2019)Huppmann, Gidden, Fricko, Kolp, Orthofer,
  Pimmer, Kushin, Vinca, Mastrucci, Riahi, and Krey}]{MESSAGE_2019}
Huppmann, Daniel, Matthew Gidden, Oliver Fricko, Peter Kolp, Clara Orthofer,
  Michael Pimmer, Nikolay Kushin, Adriano Vinca, Alessio Mastrucci, Keywan
  Riahi, and Volker Krey. 2019.
\newblock \enquote{The {MESSAGE} {Integrated} {Assessment} {Model} and the ix
  modeling platform (ixmp): {An} open framework for integrated and
  cross-cutting analysis of energy, climate, the environment, and sustainable
  development.}
\newblock \emph{Environmental Modelling \& Software} 112:143--156.

\bibitem[{IPCC(2021)}]{IPCC2021}
IPCC. 2021.
\newblock \emph{Climate Change 2021, The Physical Science Basis}.
\newblock New York: Cambridge University Press.

\bibitem[{Irwin, Gopalakrishnan, and Randall(2016)}]{irwin_welfare_2016}
Irwin, Elena~G., Sathya Gopalakrishnan, and Alan Randall. 2016.
\newblock \enquote{Welfare, {Wealth}, and {Sustainability}.}
\newblock \emph{Annual Review of Resource Economics} 8~(1):77--98.

\bibitem[{Iverson(2012)}]{Iverson2012}
Iverson, Terrence. 2012.
\newblock \enquote{Communicating Trade-offs amid Controversial Science:
  Decision Support for Climate Policy.}
\newblock \emph{Ecological Economics} 77:74--90.

\bibitem[{Iverson and Karp(2021)}]{iverson2021carbon}
Iverson, Terrence and Larry Karp. 2021.
\newblock \enquote{Carbon taxes and climate commitment with non-constant time
  preference.}
\newblock \emph{The Review of economic studies} 88~(2):764--799.

\bibitem[{Johnson et~al.(2025)Johnson, Chaplin-Kramer, Chapman, Polasky, and
  Williams}]{Johnson2025}
Johnson, Justin~Andrew, Rebecca Chaplin-Kramer, Melissa Chapman, Stephen
  Polasky, and Brooke Williams. 2025.
\newblock \enquote{Earth-Economy Modeling: Advances in Linking Economic and
  Ecosystem Models.}
\newblock \emph{Annual Review of Resource Economics} 17:209--239.

\bibitem[{Kelly et~al.(2024)Kelly, Heutel, Moreno-Cruz, and
  Shayegh}]{Kelly2024}
Kelly, David~L., Garth Heutel, Juan~B. Moreno-Cruz, and Soheil Shayegh. 2024.
\newblock \enquote{Solar Geoengineering, Learning, and Experimentation.}
\newblock \emph{Journal of the Association of Environmental and Resource
  Economists} 11~(6):1447--1486.

\bibitem[{Kelly and Kolstad(1999)}]{KellyKolstad1999}
Kelly, David~L. and Charles~D. Kolstad. 1999.
\newblock \enquote{Bayesian learning, growth, and pollution.}
\newblock \emph{Journal of Economic Dynamics and Control} 23:491--518.

\bibitem[{Kelly and Tan(2015)}]{KellyTan2015}
Kelly, David~L. and Zhuo Tan. 2015.
\newblock \enquote{Learning and climate feedbacks: optimal climate insurance
  and fat tails.}
\newblock \emph{Journal of Environmental Economics and Management} 72:98--122.

\bibitem[{Khanna et~al.(2024)Khanna, Atallah, Heckelei, Wu, and
  Storm}]{Khanna-2024}
Khanna, Madhu, Shady~S. Atallah, Thomas Heckelei, Linghui Wu, and Hugo Storm.
  2024.
\newblock \enquote{Economics of the Adoption of Artificial Intelligence-Based
  Digital Technologies in Agriculture.}
\newblock \emph{Annual Review of Resource Economics} 16:41--61.

\bibitem[{Kling et~al.(2017)Kling, Arritt, Calhoun, and Keiser}]{Kling-2017}
Kling, Catherine~L., Raymond~W. Arritt, Gray Calhoun, and David~A. Keiser.
  2017.
\newblock \enquote{Integrated Assessment Models of the Food, Energy, and Water
  Nexus: A Review and an Outline of Research Needs.}
\newblock \emph{Annual Review of Resource Economics} 9:143--163.

\bibitem[{Kotlikoff et~al.(2021)Kotlikoff, Kubler, Polbin, Sachs, and
  Scheidegger}]{Kotlikoff-2021}
Kotlikoff, Laurence, Felix Kubler, Andrey Polbin, Jeffrey Sachs, and Simon
  Scheidegger. 2021.
\newblock \enquote{Making Carbon Taxation a Generational Win Win.}
\newblock \emph{International Economic Review} 62~(1):3--46.

\bibitem[{Kotlikoff et~al.(2024)Kotlikoff, Kubler, Polbin, and
  Scheidegger}]{KOTLIKOFF2024}
Kotlikoff, Laurence, Felix Kubler, Andrey Polbin, and Simon Scheidegger. 2024.
\newblock \enquote{Can today's and tomorrow's world uniformly gain from carbon
  taxation?}
\newblock \emph{European Economic Review} 168:104819.

\bibitem[{Koundouri et~al.(2025)Koundouri, Landis, Dellis, and
  Plataniotis}]{Koundouri2025}
Koundouri, Phoebe, Conrad Felix~Michel Landis, Konstantinos Dellis, and Angelos
  Plataniotis. 2025.
\newblock \enquote{Integrating Sustainable Development Goals in Environmental,
  Social, and Governance Criteria and the Sustainability Transformation of the
  EU Business Sector.}
\newblock \emph{Annual Review of Resource Economics} 17:381--399.

\bibitem[{Lemoine and Rudik(2017)}]{LemoineRudik2017}
Lemoine, Derek and Ivan Rudik. 2017.
\newblock \enquote{Managing Climate Change Under Uncertainty: Recursive
  Integrated Assessment at an Inflection Point.}
\newblock \emph{Annual Review of Resource Economics} 9~(1):117--142.

\bibitem[{Lemoine and Traeger(2014)}]{LemoineTraeger2014}
Lemoine, Derek and Christian Traeger. 2014.
\newblock \enquote{Watch Your Step: Optimal Policy in a Tipping Climate.}
\newblock \emph{American Economic Journal: Economic Policy} 6~(1):137--166.

\bibitem[{Liu et~al.(2015)Liu, Mooney, Hull, Davis, Gaskell, Hertel, Lubchenco,
  Seto, Gleick, Kremen, and Li}]{Liu_science_2015}
Liu, Jianguo, Harold Mooney, Vanessa Hull, Steven~J. Davis, Joanne Gaskell,
  Thomas Hertel, Jane Lubchenco, Karen~C. Seto, Peter Gleick, Claire Kremen,
  and Shuxin Li. 2015.
\newblock \enquote{Systems integration for global sustainability.}
\newblock \emph{Science} 347~(6225):1258832.

\bibitem[{Lontzek et~al.(2015)Lontzek, Cai, Judd, and
  Lenton}]{lontzek_NCC_2015}
Lontzek, Thomas~S., Yongyang Cai, Kenneth~L. Judd, and Timothy~M. Lenton. 2015.
\newblock \enquote{Stochastic integrated assessment of climate tipping points
  indicates the need for strict climate policy.}
\newblock \emph{Nature Climate Change} 5~(5):441--444.

\bibitem[{Luderer et~al.(2015)Luderer, Leimbach, Bauer, Kriegler, Baumstark,
  Bertram, Giannousakis, Hilaire, Klein, Levesque, Mouratiadou, Pehl,
  Pietzcker, Piontek, Roming, Schultes, Schwanitz, and Strefler}]{REMIND_2015}
Luderer, Gunnar, Marian Leimbach, Nico Bauer, Elmar Kriegler, Lavinia
  Baumstark, Christoph Bertram, Anastasis Giannousakis, Jerome Hilaire, David
  Klein, Antoine Levesque, Ioanna Mouratiadou, Michaja Pehl, Robert Pietzcker,
  Franziska Piontek, Niklas Roming, Anselm Schultes, Valeria~Jana Schwanitz,
  and Jessica Strefler. 2015.
\newblock \enquote{Description of the {REMIND} {Model} ({Version} 1.6).}
\newblock {SSRN} {Scholarly} {Paper} ID 2697070.
\newblock \urlprefix\url{https://papers.ssrn.com/abstract=2697070}.

\bibitem[{Manne, Mendelsohn, and Richels(1995)}]{manne_merge_1995}
Manne, Alan, Robert Mendelsohn, and Richard Richels. 1995.
\newblock \enquote{{MERGE}: {A} model for evaluating regional and global
  effects of {GHG} reduction policies.}
\newblock \emph{Energy Policy} 23~(1):17--34.

\bibitem[{Manne and Richels(2005)}]{Manne_merge_2005}
Manne, Alan~S. and Richard~G. Richels. 2005.
\newblock \enquote{{MERGE}: {An} {Integrated} {Assessment} {Model} for {Global}
  {Climate} {Change}.}
\newblock In \emph{Energy and {Environment}}, edited by Richard Loulou,
  Jean-Philippe Waaub, and Georges Zaccour. New York: Springer-Verlag,
  175--189.

\bibitem[{Matthews et~al.(2009)Matthews, Gillett, Stott, and
  Zickfeld}]{matthews_proportionality_2009}
Matthews, H.~Damon, Nathan~P. Gillett, Peter~A. Stott, and Kirsten Zickfeld.
  2009.
\newblock \enquote{The proportionality of global warming to cumulative carbon
  emissions.}
\newblock \emph{Nature} 459~(7248):829--832.

\bibitem[{McKay et~al.(2022)McKay, Staal, Abrams, Winkelmann, Sakschewski,
  Loriani, Fetzer, Cornell, Rockstrom, and
  Lenton}]{Armstrong_McKay_Science2022}
McKay, David I.~Armstrong, Arie Staal, Jesse~F. Abrams, Ricarda Winkelmann,
  Boris Sakschewski, Sina Loriani, Ingo Fetzer, Sarah~E. Cornell, Johan
  Rockstrom, and Timothy~M. Lenton. 2022.
\newblock \enquote{Exceeding 1.5C global warming could trigger multiple climate
  tipping points.}
\newblock \emph{Science} 377~(6611):eabn7950.

\bibitem[{Meinshausen et~al.(2011)Meinshausen, Smith, Calvin, Daniel, Kainuma,
  Lamarque, Matsumoto, Montzka, Raper, Riahi, Thomson, Velders, and van
  Vuuren}]{Meinshausen_RCP}
Meinshausen, M., S.J. Smith, K.~Calvin, J.S. Daniel, M.L.T. Kainuma, J-F.
  Lamarque, K.~Matsumoto, S.A. Montzka, S.C.B. Raper, K.~Riahi, A.~Thomson,
  G.J.M. Velders, and D.P.P. van Vuuren. 2011.
\newblock \enquote{The {RCP} greenhouse gas concentrations and their extensions
  from 1765 to 2300.}
\newblock \emph{Climatic Change} 109:213--241.

\bibitem[{Miao and Khanna(2020)}]{Miao-2020}
Miao, Ruiqing and Madhu Khanna. 2020.
\newblock \enquote{Harnessing Advances in Agricultural Technologies to Optimize
  Resource Utilization in the Food-Energy-Water Nexus.}
\newblock \emph{Annual Review of Resource Economics} 12:65--85.

\bibitem[{Millar et~al.(2017)Millar, Nicholls, Friedlingstein, and
  Allen}]{millar_FAIR_2017}
Millar, R.~J., Z.~R. Nicholls, P.~Friedlingstein, and M.~R. Allen. 2017.
\newblock \enquote{A modified impulse-response representation of the global
  near-surface air temperature and atmospheric concentration response to carbon
  dioxide emissions.}
\newblock \emph{Atmospheric Chemistry and Physics} 17~(11):7213--7228.

\bibitem[{Navarro-Racines et~al.(2020)Navarro-Racines, Tarapues, Thornton,
  Jarvis, and Ramirez-Villegas}]{navarro-racines_2020}
Navarro-Racines, Carlos, Jaime Tarapues, Philip Thornton, Andy Jarvis, and
  Julian Ramirez-Villegas. 2020.
\newblock \enquote{High-resolution and bias-corrected {CMIP5} projections for
  climate change impact assessments.}
\newblock \emph{Scientific Data} 7~(1):7.

\bibitem[{Nordhaus(1992)}]{Nordhaus1992}
Nordhaus, William~D. 1992.
\newblock \enquote{An Optimal Transition Path for Controlling Greenhouse
  Gases.}
\newblock \emph{Science} 258~(5086):1315--1319.

\bibitem[{Nordhaus(2008)}]{Nordhaus_DICE2007}
---{}---{}---. 2008.
\newblock \emph{A Question of Balance: Weighing the Options on Global Warming
  Policies}.
\newblock Yale University Press.

\bibitem[{Nordhaus(2010)}]{Nordhaus_RICE_2010}
---{}---{}---. 2010.
\newblock \enquote{Economic aspects of global warming in a post-Copenhagen
  environment.}
\newblock \emph{Proceedings of the National Academy of Sciences}
  107~(26):11721--11726.

\bibitem[{Nordhaus(2017)}]{Nordhaus_DICE2016}
---{}---{}---. 2017.
\newblock \enquote{Revisiting the social cost of carbon.}
\newblock \emph{Proceedings of the National Academy of Sciences of the United
  States of America} 114~(7):1518--1523.

\bibitem[{Nordhaus(2019)}]{nordhaus_economics_2019}
---{}---{}---. 2019.
\newblock \enquote{Economics of the disintegration of the {Greenland} ice
  sheet.}
\newblock \emph{Proceedings of the National Academy of Sciences}
  116~(25):12261--12269.

\bibitem[{Nordhaus and Yang(1996)}]{nordhaus_yang_1996}
Nordhaus, William~D. and Zili Yang. 1996.
\newblock \enquote{A Regional Dynamic General Equilibrium Model of Alternative
  Climate Change Strategies.}
\newblock \emph{The American Economic Review} 86~(4):741--765.

\bibitem[{Olijslagers and van Wijnbergen(2024)}]{Olijslagers2024}
Olijslagers, S. and S.~van Wijnbergen. 2024.
\newblock \enquote{Discounting the Future: On Climate Change, Ambiguity
  Aversion and Epstein-Zin Preferences.}
\newblock \emph{Environ Resource Econ} 87:683--730.

\bibitem[{O'Neill et~al.(2014)O'Neill, Kriegler, Riahi, Ebi, Hallegatte,
  Carter, Mathur, and van Vuuren}]{ONeill_etal2014}
O'Neill, Brian~C., Elmar Kriegler, Keywan Riahi, Kristie~L. Ebi, Stephane
  Hallegatte, Timothy~R. Carter, Ritu Mathur, and Detlef~P. van Vuuren. 2014.
\newblock \enquote{A new scenario framework for climate change research: the
  concept of shared socioeconomic pathways.}
\newblock \emph{Climatic Change} 122~(3):387--400.

\bibitem[{Paudel(2025)}]{Paudel2025}
Paudel, Jayash. 2025.
\newblock \enquote{Effects of Natural Disasters on Human Capital.}
\newblock \emph{Annual Review of Resource Economics} 17~(Volume 17,
  2025):323--337.

\bibitem[{Riahi et~al.(2017)Riahi, {van Vuuren}, Kriegler, Edmonds, O'Neill,
  Fujimori, Bauer, Calvin, Dellink, Fricko, Lutz, Popp, Cuaresma, KC, Leimbach,
  Jiang, Kram, Rao, Emmerling, Ebi, Hasegawa, Havlik, Humpenoder, {Da Silva},
  Smith, Stehfest, Bosetti, Eom, Gernaat, Masui, Rogelj, Strefler, Drouet,
  Krey, Luderer, Harmsen, Takahashi, Baumstark, Doelman, Kainuma, Klimont,
  Marangoni, Lotze-Campen, Obersteiner, Tabeau, and Tavoni}]{RIAHI2017}
Riahi, Keywan, Detlef~P. {van Vuuren}, Elmar Kriegler, Jae Edmonds, Brian~C.
  O'Neill, Shinichiro Fujimori, Nico Bauer, Katherine Calvin, Rob Dellink,
  Oliver Fricko, Wolfgang Lutz, Alexander Popp, Jesus~Crespo Cuaresma, Samir
  KC, Marian Leimbach, Leiwen Jiang, Tom Kram, Shilpa Rao, Johannes Emmerling,
  Kristie Ebi, Tomoko Hasegawa, Petr Havlik, Florian Humpenoder, Lara~Aleluia
  {Da Silva}, Steve Smith, Elke Stehfest, Valentina Bosetti, Jiyong Eom, David
  Gernaat, Toshihiko Masui, Joeri Rogelj, Jessica Strefler, Laurent Drouet,
  Volker Krey, Gunnar Luderer, Mathijs Harmsen, Kiyoshi Takahashi, Lavinia
  Baumstark, Jonathan~C. Doelman, Mikiko Kainuma, Zbigniew Klimont, Giacomo
  Marangoni, Hermann Lotze-Campen, Michael Obersteiner, Andrzej Tabeau, and
  Massimo Tavoni. 2017.
\newblock \enquote{The Shared Socioeconomic Pathways and their energy, land
  use, and greenhouse gas emissions implications: An overview.}
\newblock \emph{Global Environmental Change} 42:153--168.

\bibitem[{Rudik(2020)}]{Rudik_2020}
Rudik, Ivan. 2020.
\newblock \enquote{Optimal Climate Policy When Damages Are Unknown.}
\newblock \emph{American Economic Journal: Economic Policy} 12~(2):340--373.

\bibitem[{Sohngen(2020)}]{Brent-2020}
Sohngen, Brent. 2020.
\newblock \enquote{Climate Change and Forests.}
\newblock \emph{Annual Review of Resource Economics} 12:23--43.

\bibitem[{Stehfest et~al.(2014)Stehfest, van Vuuren, Kram, Bouwman, Alkemade,
  Bakkenes, Biemans, Bouwman, den Elzen, Janse, Lucas, van Minnen, M¸ller, and
  Prins}]{IMAGE_2014}
Stehfest, E., D.~van Vuuren, T.~Kram, L.~Bouwman, R.~Alkemade, M.~Bakkenes,
  H.~Biemans, A.~Bouwman, M.~den Elzen, J.~Janse, P.~Lucas, J.~van Minnen,
  C.~M¸ller, and A.~Prins. 2014.
\newblock \emph{Integrated Assessment of Global Environmental Change with
  {IMAGE} 3.0 - Model description and policy applications}.
\newblock PBL Netherlands Environmental Assessment Agency.

\bibitem[{Steinbuks et~al.(2024)Steinbuks, Cai, Jaegermeyr, and
  Hertel}]{Steinbuks-2024}
Steinbuks, Jevgenijs, Yongyang Cai, Jonas Jaegermeyr, and Thomas~W. Hertel.
  2024.
\newblock \enquote{Assessing effects of climate and technology uncertainties in
  large natural resource allocation problems.}
\newblock \emph{Geoscientific Model Development} 17~(12):4791--4819.

\bibitem[{Stern(2007)}]{Stern2007}
Stern, N.~H. 2007.
\newblock \emph{The economics of climate change: the Stern Review}.
\newblock Cambridge, UK: Cambridge University Press.

\bibitem[{Tibebu et~al.(2025)Tibebu, Li, Torres~Arroyo, Lessard, Bozeman~III,
  Cai, Gephart, Konar, Lee, Romeiko, Talley, and Siddiqui}]{Tibebu_2025}
Tibebu, Tiruwork~B, Siyu Li, Mariana Torres~Arroyo, Katherine Lessard, Joe~F
  Bozeman~III, Yongyang Cai, Jessica~A Gephart, Megan Konar, Young-Jae Lee,
  Xiaobo Romeiko, Jessye Talley, and Sauleh Siddiqui. 2025.
\newblock \enquote{Interactions and tradeoffs for sustainability, equity, and
  resilience in wasted food models.}
\newblock \emph{Environmental Research Communications} 7~(4):045013.

\bibitem[{Tol(1997)}]{Tol1997}
Tol, Richard. 1997.
\newblock \enquote{On the optimal control of carbon dioxide emissions: an
  application of {FUND}.}
\newblock \emph{Environmental Modeling and Assessment} 2:151--163.

\bibitem[{van~den Bremer and van~der Ploeg(2021)}]{Bremer_AER2021}
van~den Bremer, Ton~S. and Frederick van~der Ploeg. 2021.
\newblock \enquote{The Risk-Adjusted Carbon Price.}
\newblock \emph{American Economic Review} 111~(9):2782--2810.

\bibitem[{van~der Ploeg and de~Zeeuw(2018)}]{van_der_ploeg_climate_2018}
van~der Ploeg, Frederick and Aart de~Zeeuw. 2018.
\newblock \enquote{Climate {Tipping} and {Economic} {Growth}: {Precautionary}
  {Capital} and the {Price} of {Carbon}.}
\newblock \emph{Journal of the European Economic Association}
  16~(5):1577--1617.

\bibitem[{van~der Ploeg and de~Zeeuw(2019)}]{van_der_ploeg_pricing_2019}
---{}---{}---. 2019.
\newblock \enquote{Pricing {Carbon} and {Adjusting} {Capital} to {Fend} {Off}
  {Climate} {Catastrophes}.}
\newblock \emph{Environmental and Resource Economics} 72~(1):29--50.

\bibitem[{van~der Ploeg and Rezai(2020)}]{van_der_ploeg_risk_2020}
van~der Ploeg, Frederick and Armon Rezai. 2020.
\newblock \enquote{The risk of policy tipping and stranded carbon assets.}
\newblock \emph{Journal of Environmental Economics and Management} 100~(Article
  102258).

\bibitem[{Weitzman(2012)}]{Weitzman2012}
Weitzman, Martin~L. 2012.
\newblock \enquote{GHG Targets as Insurance Against Catastrophic Climate
  Damages.}
\newblock \emph{Journal of Public Economic Theory} 14~(2):221--244.

\bibitem[{Yang(2023)}]{yang_model_2023}
Yang, Zili. 2023.
\newblock \enquote{The {Model} {Dimensionality} and {Its} {Impacts} on the
  {Strategic} and {Policy} {Outcomes} in {IAMs}: the {Findings} from the
  {RICE2020} {Model}.}
\newblock \emph{Computational Economics} 62~(3):1087--1106.

\bibitem[{Zhao et~al.(2023)Zhao, Basu, Lontzek, and Schmedders}]{Zhao_MS_2023}
Zhao, Yifan, Arnab Basu, Thomas~S. Lontzek, and Karl Schmedders. 2023.
\newblock \enquote{The Social Cost of Carbon When We Wish for Full-Path
  Robustness.}
\newblock \emph{Management Science} 69~(12):7585--7606.

\bibitem[{Zhu et~al.(2025)Zhu, Yan, Duan, Cai, and Zhang}]{Zhu_2025}
Zhu, Lei, Zhihao Yan, Hongbo Duan, Yongyang Cai, and Xiaobing Zhang. 2025.
\newblock \enquote{Long Coalition Leads to Shrink? The Roles of Tipping and
  Technology-Sharing in Climate Clubs.}
\newblock arXiv Working Paper 2506.16162.
\newblock \urlprefix\url{https://doi.org/10.48550/arXiv.2506.16162}.

\end{thebibliography}

\end{document}